\pdfoutput=1
\documentclass[%
superscriptaddress,
nofootinbib,
twocolumn,
amsmath,amssymb,
aps,
prd
]{revtex4-2}

\usepackage{graphicx}
\usepackage{dcolumn}
\usepackage{bm}
\usepackage{color}

\def\araa{{ARA\&A}}             
\def\apj{{ApJ}}                 
\def\apss{{Ap\&SS}}             
\def\aap{{A\&A}}                

\def\jcap{{J. Cosmology Astropart. Phys.}}
\def\mnras{Mon.~Not.~Roy.~Astron.~Soc.}             
\def\prd{{Phys.~Rev.~D}}        
\def\physrep{{Phys.~Rep.}}   

\newcommand{\beq}{\begin{equation}}
\newcommand{\beqa}{\begin{eqnarray}}
\newcommand{\eeq}{\end{equation}}
\newcommand{\eeqa}{\end{eqnarray}}

\newcommand{\simgt}{\lower.5ex\hbox{$\; \buildrel > \over \sim \;$}}
\newcommand{\simlt}{\lower.5ex\hbox{$\; \buildrel < \over \sim \;$}}

\newcommand{\bd}[1]{\mbox{\boldmath $#1$}}
\definecolor{armygreen}{rgb}{0.29, 0.33, 0.13}
\newcommand{\rev}[1]{\textcolor{black}{#1}}

\usepackage[usenames,dvipsnames]{xcolor}

\begin{document}


\title{Cross Correlation of the Extragalactic Gamma-ray Background with \\
Thermal Sunyaev-Zel'dovich Effect in the Cosmic Microwave Background
}

\author{Masato Shirasaki}
\email{masato.shirasaki@nao.ac.jp}
\affiliation{
National Astronomical Observatory of Japan (NAOJ), 
Mitaka, Tokyo 181-8588, Japan}

\author{Oscar Macias}
\email{oscar.macias@ipmu.jp}
\affiliation{
Kavli Institute for the Physics and Mathematics of the Universe (WPI), 
University of Tokyo, Kashiwa, Chiba 277-8583, Japan}
\affiliation{
GRAPPA Institute, University of Amsterdam, 1098 XH Amsterdam, The Netherlands}
\author{Shin'ichiro Ando}
\email{s.ando@uva.nl}
\affiliation{
GRAPPA Institute, University of Amsterdam, 1098 XH Amsterdam, The Netherlands}
\affiliation{
Kavli Institute for the Physics and Mathematics of the Universe (WPI), 
University of Tokyo, Kashiwa, Chiba 277-8583, Japan}
\author{Shunsaku Horiuchi}
\email{horiuchi@vt.edu}
\affiliation{Center for Neutrino Physics, Department of Physics,Virginia Tech, Blacksburg, Virginia 24061, USA}
\author{Naoki Yoshida}
\affiliation{
 Department of Physics, University of Tokyo, Tokyo 113-0033, Japan
}%
\affiliation{
Kavli Institute for the Physics and Mathematics of the Universe (WPI), 
University of Tokyo, Kashiwa, Chiba 277-8583, Japan}

\affiliation{
Institute for Physics of Intelligence, University of Tokyo, Tokyo 113-0033, Japan
}
\affiliation{
Research Center for the Early Universe, Faculty of Science, University of Tokyo, Tokyo 113-0033, Japan
}
\date{\today}

\begin{abstract}
Cosmic rays in galaxy clusters are unique probes of energetic processes 
operating with large-scale structures in the Universe. Precise measurements of cosmic rays 
in galaxy clusters are essential for improving our understanding of non-thermal components 
in the intracluster-medium (ICM) as well as the accuracy of cluster mass estimates in cosmological analyses. 
In this paper, we perform a cross-correlation analysis with the extragalactic gamma-ray background 
and the thermal Sunyaev-Zel’dovich (tSZ) effect in the cosmic microwave background. 
The expected cross-correlation signal would contain rich information about the cosmic-ray-induced gamma-ray emission in the most massive galaxy clusters at $z\sim0.1$--0.2. 
We analyze the gamma-ray background map with 8 years of data 
taken by the Large Area Telescope onboard Fermi satellite and the publicly available tSZ map by Planck. 
We confirm that the measured cross-correlation is consistent with a null detection, and thus it enables us to 
put the tightest constraint on the acceleration efficiency of cosmic ray protons at shocks in and around galaxy clusters.
We find the acceleration efficiency must be below 5\% with a $2\sigma$ confidence level when 
the hydrostatic mass bias of clusters is assumed to be 30\%,
and our result is not significantly affected by the assumed value of the hydrostatic mass bias.
Our constraint implies that the non-thermal cosmic-ray pressure in the ICM can introduce only a $\simlt 3\%$
level of the hydrostatic mass bias, highlighting that cosmic rays alone 
do not account for
the mass bias inferred by the Planck analyses. 
Finally, we discuss future detectability prospects of cosmic-ray-induced gamma rays from the Perseus cluster 
for the Cherenkov Telescope Array. 
\end{abstract}

\maketitle

\section{\label{sec:intro}INTRODUCTION}


Galaxy clusters are known to be the most massive self-bound objects in the Universe. The standard structure formation
theory predicts that galaxy clusters form through a hierarchical
sequence of mergers and accretion of smaller objects driven 
by the gravitational growth of cosmic mass density \cite{2012ARA&A..50..353K}.
Mergers are one of the most energetic phenomena in the Universe, generating shocks around galaxy clusters and 
heating the gas temperature 
in the intra-cluster medium (ICM).
Detailed studies of dissipation of 
the gravitational energy in the cluster formation
will be key to understanding the nature of the ICM. This is because the processes of dissipation can cause the production
of non-thermal components in the ICM, 
such as relativistic particles, or cosmic rays \cite{Brunetti:2014gsa}.
Understanding the ICM physics 
enables us to estimate the masses of individual clusters 
from multi-wavelength observations accurately
and perform precise cosmological analyses 
based on the cluster number count \cite{2011ARA&A..49..409A}.


Radio observations of galaxy clusters have found 
diffuse synchrotron radiation from the ICM \cite{Ferrari:2008jr}. 
The detected synchrotron radiation from galaxy clusters provides the main evidence for large-scale magnetic fields and cosmic-ray electrons in the ICM.
As a natural consequence, galaxy clusters should 
confine cosmic-ray protons (hadrons) over cosmological times
because of the long lifetime of cosmic-ray protons against energy losses and the slow diffusive propagation in the ICM magnetic fields.
The detection of gamma-ray emission 
produced by the decay of secondary $\pi^0$ particles is 
the most direct probe of cosmic-ray protons 
in galaxy clusters.
Despite intense efforts in gamma-ray astronomy, 
no conclusive evidence for gamma-ray emission in the ICM 
has been reported so far \cite{2009A&A...502..437A, Ackermann:2010qj, 2012JCAP...07..017A, Pfrommer:2012mm, 2012MNRAS.427.1651H, Huber:2013cia, Ackermann:2013iaq, Zandanel:2013wea, Rephaeli:2015nca, Ahnen:2016qkt} (but see Ref.~\cite{Xi:2017uzz} for the recent update).


Most previous searches for gamma-ray emission 
from the ICM rely on targeted observations of single nearby galaxy clusters and suffer from limited statistics.
For a complementary approach to the previous ones, 
we propose a cross-correlation analysis of the unresolved extragalactic gamma-ray background (UGRB)
with the thermal Sunyaev-Zel'dovich (tSZ) effect 
in the comic microwave background (CMB).
The tSZ effect is known as the frequency-dependent distortion in the CMB intensity induced by the inverse Compton 
scattering of the CMB photons in the ICM \cite{1969Ap&SS...4..301Z, 1972CoASP...4..173S}.
The Planck satellite has completed its survey operation over about four years \cite{Akrami:2018vks}. 
The multi-frequency bands in the Planck enabled us to obtain the cleanest map of CMB so far \cite{Ade:2013crn, Adam:2015tpy, Akrami:2018mcd} and reconstruct the tSZ effect on a line-of-sight basis over a wide sky \cite{Ade:2013qta, Aghanim:2015eva}. 
Hence, the Planck tSZ map can provide a unique 
opportunity to probe the ICM without any selection effects 
of galaxy clusters. 
Since the UGRB is expected to be the cumulative emission from faint gamma ray sources, 
it may also 
contain valuable information on the ICM, if the ICM emits gamma rays.
In this paper, we perform, for the first time, 
the correlation analysis between the UGRB and the tSZ effect by using gamma-ray data from the Fermi and 
the publicly available Planck map. 
We also develop a theoretical model 
of the cross correlation based on the standard structure formation and the simulation-calibrated cosmic-ray model \cite{2010MNRAS.409..449P}.
Compared with our measurement and theoretical prediction,
we constrain the amount of cosmic-ray-induced 
gamma rays in the ICM in the energy range of $>700$ MeV,
at which the cosmic ray protons play a central role in possible gamma-ray emission.

It would be worth noting that a cross correlation between the UGRB and the number density of galaxy clusters is a similar statistical approach to search for the gamma rays from galaxy clusters \cite{Branchini:2016glc, Hashimoto:2018ztv, Colavincenzo:2019jtj}. This number-density-based method 
will be sensitive to the gamma-ray emission from the active Galactic nuclei (AGN) 
inside galaxy clusters, 
while our approach uses a more direct probe of the ICM 
and can provide comprehensive information about the gamma rays from the ICM.
Note that the tSZ effect mainly arises from thermal electrons in the ICM, 
while the gamma-ray emission is caused by non-thermal components. 
Hence, the cross correlation between UGRB and tSZ maps may not have the strict same origin, but
signals should be interpreted as a spatial correlation.

The paper is organized as follows. 
In Section~\ref{sec:obs}, 
we summarize the basics of UGRB
and the tSZ effect. 
Our benchmark model of
the cross correlation is discussed 
in Section~\ref{sec:model}.
In Section~\ref{sec:data}, we describe 
the gamma-ray and the tSZ data used, 
and provide details of the cross-correlation analysis. 
In Section~\ref{sec:res}, we show the result of 
our cross-correlation analysis, and 
discuss constraints on the gamma rays in the ICM.
Concluding remarks and discussions are given in Section~\ref{sec:con}. 
Throughout, we use the standard cosmological parameters
$H_0=100h\, {\rm km}\, {\rm s}^{-1}$
with $h=0.68$, the average matter density 
$\Omega_{\rm m0}=0.315$, 
the cosmological constant $\Omega_{\Lambda}=0.685$,
and the amplitude of matter density fluctuations within 
$8\,h^{-1}\, {\rm Mpc}$, $\sigma_8=0.83$.

\section{OBSERVABLES}
\label{sec:obs}

\subsection{Extragalactic gamma-ray background}
\label{subsec:EGB}

The gamma-ray intensity $I_\gamma$ is defined by 
the number of photons per unit energy, area, time, and solid angle,
\begin{equation}\label{eq:Intensity}
E_\gamma I_\gamma = \frac{c}{4\pi} \int {\rm d}z \frac{P_\gamma (E'_\gamma,z)}{H(z)(1+z)^4} e^{-\tau(E'_\gamma,z)},
\end{equation}
where $E_\gamma$ is the observed gamma-ray energy, 
$E'_\gamma = (1+z) E_\gamma$ is the energy 
of the gamma ray at redshift $z$, 
$H(z) = H_0 [\Omega_{\rm m0}(1+z)^3+\Omega_\Lambda]^{1/2}$ 
is the Hubble parameter in a flat Universe, 
and the exponential factor in the integral
takes into account the effect of gamma-ray attenuation 
during propagation owing to pair creation 
on diffuse extragalactic photons. 
For the gamma-ray optical depth $\tau\left(E'_\gamma, z \right)$, 
we adopt the model in Ref.~\citep{Gilmore:2011ks}.
Ref.~\cite{Ajello:2018sxm} has shown that 
the model in Ref.~\citep{Gilmore:2011ks} can provide 
a reasonable fit to 
the gamma-ray attenuation in the energy spectra of blazars and a gamma-ray burst.
In Eq.~(\ref{eq:Intensity}), $P_\gamma$ represents 
the volume emissivity 
(i.e., the photon energy emitted per unit volume, time, and energy range),
which is given by
\beqa\label{eq:dmEmissivity}
P_\gamma(E_\gamma, z)=
E_\gamma {\cal S}(E_\gamma, z) 
{\cal F}(\bd{r}, z),
\eeqa
where ${\cal S}$ 
is a gamma-ray source function and ${\cal F}$ represents 
the relevant density field of gamma-ray sources.

In this paper, we assume that the UGRB intensity is measured in the energy range
$E_{\gamma, {\rm min}}$ to $E_{\gamma, {\rm max}}$ 
along a given angular direction $\hat{\bd n}$.
In this case, the more relevant formula is given by 
\beqa
I_{\gamma}(\hat{\bd n}) &=& \int {\rm d}\chi
\, W_{\gamma}(\chi) {\cal F}(\chi \hat{\bd n}, z(\chi)), 
\label{eq:Intensity_dm}
\\
W_{\gamma}(\chi) &=&
\int_{E_{\gamma, {\rm min}}}^{E_{\gamma, {\rm max}}}
\frac{{\rm d}E_{\gamma}}{4\pi} \, \frac{{\cal S}(E'_{\gamma}, z(\chi))}{(1+z(\chi))^3} 
e^{-\tau(E'_\gamma, z(\chi))}, \label{eq:gamma_kernel}
\eeqa
where $\chi(z)$ is the comoving distance.
In practice, we need to take into account the smearing effect in a map 
due to the point spread function (PSF) in gamma-ray measurements. 
In this paper, we apply the same framework to include this PSF effect as in Ref.~\cite{Shirasaki:2014noa},
while we update the parameters in the PSF to follow the latest Fermi pipeline accordingly. 

\subsection{Thermal Sunyaev-Zel'dovich effect}
\label{subsec:tSZ}

The tSZ effect probes the thermal pressure of hot electrons in galaxy clusters. 
At frequency $\nu$, the change in CMB temperature by the tSZ effect is expressed as
\beq
\frac{\Delta T}{T_0} = g(x) y, \label{eq:T-y}
\eeq
where $T_{0}=2.725\, {\rm K}$ is the CMB temperature \citep{2009ApJ...707..916F}, 
$g(x) = x{\rm coth}(x/2)-4$ with $x=h_{\rm P}\nu/k_{\rm B}T_{0}$,
$h_{\rm P}$ and $k_{\rm B}$ are the Planck constant and the Boltzmann constant, respectively\footnote{In this paper, we ignore the relativistic correction for $g(x)$ which is a secondary effect in the current tSZ measurements \cite{1998ApJ...502....7I,2012MNRAS.426..510C}.}.
Compton parameter $y$ is obtained as the integral of the electron pressure $P_{\rm e}$ along a line of sight:
\beq
y(\hat{\bd n}) = \int\, \frac{{\rm d}\chi}{1+z}\, \frac{k_{\rm B}\sigma_{\rm T}}{m_{\rm e}c^2} P_{\rm e}\left(\chi\hat{\bd n}, z(\chi)\right), \label{eq:tSZ_y}
\eeq
where $\sigma_{\rm T}$ is the Thomson cross section.

\section{\label{sec:model}ANALYTIC MODEL OF CROSS POWER SPECTRUM}

In this section, we describe the formalism to compute 
the cross power spectra between the UGRB intensity $I_\gamma$ and the tSZ Compton parameter $y$. 
The cross power spectrum between any two fields is given by: 
\beq
\langle {\cal A}({\bd \ell}_1){\cal B}({\bd \ell}_{2})\rangle \equiv (2\pi)^2 \delta^{(2)}_{\rm D}({\bd \ell}_1-{\bd \ell}_2)
C_{{\cal A}{\cal B}}(\ell_{1}), 
\eeq
where $\langle \cdots \rangle$ indicates the operation of ensemble average,
$\delta^{(n)}_{\rm D}({\bd r})$ represents the Dirac delta function in $n$-dimensional space,
$\cal A$ and $\cal B$ are projected fields of interest.

\subsection{Halo Model Approach}
The cross power spectra for any two fields $C_{{\cal A}{\cal B}}$, 
under the flat-sky approximation\footnote{The exact expression for the curved sky can be found in Appendix~A of Ref.~\cite{hill13}.}, can be decomposed into two components within the halo-model framework \cite{2002PhR...372....1C} 
\beq
C_{{\cal A}{\cal B}}(\ell) = C^{\rm 1h}_{{\cal A}{\cal B}}(\ell) + C^{\rm 2h}_{{\cal A}{\cal B}}(\ell), \label{eq:halo_model_power_ab}
\eeq
where the first term on the right-hand side represents the two-point clustering in a single halo (i.e. the 1-halo term),
and the second corresponds to the clustering term between a pair of halos (i.e. the 2-halo term).
They are expressed as 
\cite{1999ApJ...526L...1K, komatsu02, hill13}
\begin{align}
C^{\rm 1h}_{{\cal A}{\cal B}}(\ell)
=& \int_{z_{\rm min}}^{z_{\rm max}}\, {\rm d}z\, \frac{{\rm d}^2V}{{\rm d}z{\rm d}\Omega}
\int_{M_{\rm min}}^{M_{\rm max}}\, {\rm d}M\, \frac{{\rm d}n}{{\rm d}M}\, \nonumber \\
& \qquad \qquad \qquad \qquad \times |{\cal A}_{\ell}(M,z) {\cal B}_{\ell}(M,z)|, \label{eq:A_B_1h} \\
C^{\rm 2h}_{{\cal A}{\cal B}}(\ell)
=& \int_{z_{\rm min}}^{z_{\rm max}}\, {\rm d}z\, \frac{{\rm d}V}{{\rm d}z
{\rm d}\Omega}\, P_{\rm L}(k=\ell/\chi,z) \nonumber \\
& \qquad \times \left[
\int_{M_{\rm min}}^{M_{\rm max}}\, {\rm d}M\, \frac{{\rm d}n}{{\rm d}M}\, {\cal A}_{\ell}(M,z)\, b(M,z)\right] \nonumber \\
& \qquad \times \left[
\int_{M_{\rm min}}^{M_{\rm max}}\, {\rm d}M\, \frac{{\rm d}n}{{\rm d}M}\, {\cal B}_{\ell}(M,z)\, b(M,z)
\right], \label{eq:A_B_2h}
\end{align}
where we adopt $z_{\rm min}=0.01$, 
$z_{\rm max}=3$, $M_{\rm min}=10^{13}\, h^{-1}M_{\odot}$ and $M_{\rm max}=10^{16}\, h^{-1}M_{\odot}$,
$P_{\rm L}(k,z)$ is the linear matter power spectrum, ${\rm d}n/{\rm d}M$ is the halo mass function, and $b$ is the linear halo bias. We define the halo mass $M$ by virial overdensity \cite{1998ApJ...495...80B}. 
We set the minimum redshift $z_{\rm min}=0.01$ in our halo-model calculations, because it is the lowest redshift in the galaxy cluster catalog provided by the Planck \cite{Ade:2015gva}.
We adopt the simulation-calibrated halo mass function presented in Ref.~\cite{2008ApJ...688..709T} and linear bias in Ref.~\cite{2010ApJ...724..878T}.
In Eqs.~(\ref{eq:A_B_1h}) and (\ref{eq:A_B_2h}), ${\cal A}_{\ell}(z, M)$ and ${\cal B}_{\ell}(z, M)$ represent the Fourier transforms of profiles of fields ${\cal A}$ and ${\cal B}$ of a single halo 
with mass of $M$ at redshift $z$, respectively.

\subsection{ICM profiles in a single halo}
\label{subsec:ICM_single_halo}
\subsubsection{Gamma rays from pion decays}

The high-resolution hydrodynamical simulation of galaxy clusters have shown that 
the emission coming from pion decays dominates over the inverse Compton emission of 
both primary and secondary electrons for gamma rays with energies above 100 MeV \cite{2010MNRAS.409..449P}.
Hence, we assume that the ICM contribution to the UGRB intensity can be approximated 
by the cumulative gamma-ray emission arising from pion decays in single galaxy clusters.
For the gamma-ray source function ${\cal S}(E_{\gamma}, z)$, 
we use a universal model of the cosmic-ray energy spectrum in galaxy clusters developed in Ref.~\cite{2010MNRAS.409..449P}. 
For the pion-decay-induced emission in a single cluster, the relevant density profile can be expressed as \cite{2010MNRAS.409..449P}
\beqa
{\cal F}_{h}(R, M, z) = A_{\gamma}\, C_{\gamma}(R, M) \, \frac{\rho^2_{\rm gas}(R, M, z)}{\rho^2_{\rm aux}}, \label{eq:gamma_prof}
\eeqa
where $R$ is the cluster-centric radius, $C_{\gamma}(R, M)$ controls the shape of the cosmic-ray spatial distribution compared to the square of gas density profile $\rho_{\rm gas}$, 
and $A_{\gamma}$ is a dimensionless scale parameter related to the maximum cosmic-ray proton acceleration efficiency $\xi_p$ for diffusive shock acceleration\footnote{To be specific, $\xi_p$ is defined as
the maximum ratio of cosmic-ray energy density that can be injected with respect 
to the total dissipated energy at the shock.}.
In Eq.~(\ref{eq:gamma_prof}), we introduce an auxiliary variable $\rho_{\rm aux}$ so that ${\cal F}_{h}$ can be dimensionless.
Accordingly, the gamma-ray source function 
${\cal S}(E_{\gamma}, z)$ is given by
${\cal S}(E_{\gamma}, z) = 
\rho^2_{\rm aux}/(m^3_{p} c) \times {\cal G}(E_{\gamma})$,
where $m_{p}$ is the proton mass
and ${\cal G}(E_{\gamma})$ controls the shape of the gamma-ray energy spectrum. Note that ${\cal G}$ has the unit of mbarn.
See Ref.~\cite{2010MNRAS.409..449P} for the exact form of 
${\cal G}(E_{\gamma})$.
It is worth mentioning that our prediction of the cross power spectrum is independent of the exact value of $\rho_{\rm aux}$,
because the UGRB intensity in Eq.~(\ref{eq:Intensity_dm}) depends on the product of ${\cal S} \times {\cal F}_{h}$.
Besides, the presence of magnetic fields in a cluster can affect the pion-decay spectrum at 
$E_{\gamma}\simgt10^{8}$ GeV, which is well beyond our energy range of interest.

Ref.~\cite{2010MNRAS.409..449P} sets $A_{\gamma} = 1$ for $\xi_p = 0.5$ 
and $A_{\gamma}$ is expected to decrease as $\xi_{p}$ becomes smaller.
Although the $A_{\gamma}$--$\xi_{p}$ relation would be non-linear \cite{2010MNRAS.409..449P}, 
we can approximate the relation to be linear for pion-decay emission with energies $\simgt1$ GeV \cite{Zandanel:2013wea}.
In Eq.~(\ref{eq:gamma_prof}), we adopt the following functional form of $C_{\gamma}(R, M)$ 
as calibrated in Ref.~\cite{2010MNRAS.409..449P}:
\beqa
C_{\gamma}(R, M) &=& C_{\rm cen}+(C_{\rm vir}(M) - C_{\rm cen}) \nonumber\\ &&{}\times\left[1+\left(\frac{R}{R_{\rm trans}(M)}\right)^{-\beta(M)}\right]^{-1} 
,
\eeqa
where $C_{\rm cen} = 5\times10^{-7}$ and
\beqa
C_{\rm vir}(M) &=& 1.7 \times 10^{-7} \left(\frac{M_{\rm 200c}(M)}{10^{15}\, M_{\odot}}\right)^{0.51}, \label{eq:C_vir_gamma}\\
R_{\rm trans}(M) &=& 0.021 R_{\rm 200c} \left(\frac{M_{\rm 200c}(M)}{10^{15}\, M_{\odot}}\right)^{0.39}, \\
\beta(M) &=& 1.04 \left(\frac{M_{\rm 200c}(M)}{10^{15}\, M_{\odot}}\right)^{0.15},
\eeqa
where $M_{\Delta c}$ is the spherical over-density (SO) mass with respect to 
the critical density times $\Delta$ and $R_{\Delta c}$ is the SO radius\footnote{Throughout this paper,
we convert the virial mass $M$ to different SO masses $M_{\Delta c}$ as in Ref.~\cite{Hu:2002we}
assuming the mass-redshift-concentration relation in Ref.~\cite{Diemer:2014gba}.}.

For the gas density squared in Eq.~(\ref{eq:gamma_prof}), we use a generalized Navarro-Frank-White (GNFW) profile:
\beqa
\rho_{\rm gas}(R, M, z=0) &=& \frac{m_p}{X_{H} X_{e}} n_{e} (R, M)\nonumber \\
n_{e}(R, M) &=& \frac{n_{0}(M)}{x^{\beta_g} \left[1+x^{\alpha_g}\right]^{(\delta_{g}-\beta_{g})/\alpha_{g}}},
\eeqa
where $x=R/(0.2\, R_{\rm 500c})$, $\alpha_{g}=1$, $\delta_g = 2.5$,
$X_H= 0.76$ is the primordial hydrogen mass fraction, and
$X_e= 1.157$ is the ratio of electron-to-hydrogen number 
densities in the fully ionized ICM \cite{1988xrec.book.....S}.
For $z>0$, we assume the self-similar relation $\rho_{\rm gas}(z) = \rho_{\rm gas}(z=0) H^2(z)/H^2_0$ \cite{1986MNRAS.222..323K}.
We adopt the parameters $n_{0}$ and $\beta_{g}$ in Ref.~\cite{Zandanel:2013wja} in this paper.
The authors in Ref.~\cite{Zandanel:2013wja} have calibrated  
the parameters for cool-core and non-cool-core samples
with the observed tSZ and X-ray scaling relation as well as the X-ray luminosity function.
In this paper, we assume the cool-core fraction to be $f_{\rm CC}=0.5$ 
and the total gas density profile is expressed as 
$\rho_{\rm gas} = f_{\rm CC} \rho_{\rm gas, CC} + (1-f_{\rm CC}) \rho_{\rm gas, NCC}$,
where $\rho_{\rm gas, CC}$ is the gas density profile for the cool-core population and so on.

The presence of substructures in the ICM can enhance the amplitude of the gas density squared on average. This boosting effect is known as the gas clumpiness effect. When computing Eq.~(\ref{eq:gamma_prof}), we include this clumpiness effect by introducing a multiplication function as
\beqa
\rho^2_{\rm gas} &\rightarrow& C_{\rm clump} \, \rho^2_{\rm gas}, \label{eq:gas_clump}
\eeqa
where $C_{\rm clump}$ represents the gas clumpiness effect.
In this paper, we adopt the model of $C_{\rm clump}$ 
calibrated with the numerical simulation in Ref.~\cite{Battaglia:2014cga} and its form is approximated as \cite{Lakey:2019aqt}
\beqa
C_{\rm clump}(R, M) &=& 
1+\left[\frac{x}{x_{cc}(M)}\right]^{\beta_{cc}(M)} \nonumber \\
&& \times \left[1+\frac{x}{x_{cc}(M)}\right]^{\gamma_{cc}(M)-\beta_{cc}(M)},
\eeqa
where $x=R/R_{200c}$ and,
\beqa
x_{cc}(M) &=& 9.91\times10^{5}
\left(\frac{M_{200c}(M)}{10^{14}\,M_{\odot}}\right)^{-4.87}, \\
\beta_{cc}(M) &=& 0.185 
\left(\frac{M_{200c}(M)}{10^{14}\,M_{\odot}}\right)^{0.547}, \\
\gamma_{cc}(M) &=& 1.19\times10^{6}
\left(\frac{M_{200c}(M)}{10^{14}\,M_{\odot}}\right)^{-4.86}.
\eeqa

\subsubsection{Electron pressure}

When computing the Fourier counterpart of Eq.~(\ref{eq:tSZ_y}), we adopt the model of  
3D electron pressure profile in single halo $P_{{\rm e},h}$
as constrained in Ref.~\cite{2013A&A...550A.131P},
\begin{align}
P_{{\rm e}, h}(R, M, z) =& 
1.65 \times 10^{-3}\,  \left[{\rm keV}\, {\rm cm}^{-3}\right]\, E^{8/3}(z) \nonumber \\
&
\times
\left(\frac{M_{500c}(M)}{3\times10^{14}\, h_{70}^{-1}M_{\odot}}\right)^{2/3+0.12}\, {\cal P}(x)\, h_{70}^2, \label{eq:UPP_PLANCK}
\end{align}
where $x=R/R_{500c}$,
$E(z)=H(z)/H_{0}$, $h_{70}=H_0/70$, and $\mathcal{P}(x)$ is so-called universal pressure profile 
\citep{Nagai:2007mt}. The functional form of ${\cal P}(x)$ is given by
\beq
{\cal P}(x) = \frac{P_0}{(c_{500}x)^{\gamma_P}\left[1+(c_{500}x)^{\alpha_P}\right]^{(\beta_P-\gamma_P)/\alpha_P}}, \label{eq:UPP}
\eeq
where we adopt the best-fit values of five parameters 
($P_0, c_{500}, \alpha_P, \beta_P$, and $\gamma_P$) from Ref.~\cite{2013A&A...550A.131P}.
Note that the input mass parameter $M_{500c}$ in Eq.~(\ref{eq:UPP_PLANCK}) will 
be affected by hydrostatic mass bias, because the mass-scaling relation in Eq.~(\ref{eq:UPP_PLANCK}) 
has been calibrated
with the actual tSZ measurements alone.
For a given halo mass of $M$ (the virial SO mass), 
we include the hydrostatic mass bias $b_{\rm HSE}$ by $M_{500c} \rightarrow M_{500c}/(1+b_{\rm HSE})$ 
and $R_{500c} \rightarrow R_{500c}/(1+b_{\rm HSE})^{1/3}$ for Eq.~(\ref{eq:UPP_PLANCK}). 
We set $b_{\rm HSE}=0.2$ as in Ref.~\cite{Dolag:2015dta} for our baseline model.
It is worth noting that Ref.~\cite{Dolag:2015dta} shows that 
the above model of the electron pressure can explain the observed tSZ power spectrum \cite{Aghanim:2015eva}.

\subsubsection{Fourier counterparts}
\label{subsubsec:fourier_single_halo}

The Fourier transforms of the gamma-ray emissivity profile $\gamma_{\ell}(M,z)$ and the thermal electron pressure profile $y_{\ell}(M,z)$ of the halo with mass $M$ and redshift $z$ are expressed as
\beqa
\gamma_{\ell}(M,z) &=&
\frac{4\pi R_{500c}}{\ell^2_{\rm 500}}
\int\, {\rm d}u\, u^2\, \frac{\sin(\ell u/\ell_{500})}{\ell u/\ell_{500}} \nonumber \\
&& 
\qquad \qquad
\times W_{\gamma}(z, \ell) {\cal F}_{h}(uR_{500c}, M, z), \label{eq:gamma_ell} \\
y_{\ell}(M,z) &=& 
\frac{4\pi R_{500c}}{\ell^2_{\rm 500}}
\int\, {\rm d}u\, u^2\, \frac{\sin(\ell u/\ell_{500})}{\ell u/\ell_{500}} \nonumber \\
&&
\qquad \qquad
\times \frac{\sigma_{\rm T}}{m_{\rm e}c^2} P_{{\rm e},h}(uR_{500c}, M, z), \label{eq:y_ell}
\eeqa
where $u = R/R_{500c}$, $\ell_{500}=\chi/R_{500c}/(1+z)$, 
${\cal F}_{h}$ is the gamma-ray emissivity profile defined in Eq.~(\ref{eq:gamma_prof}), and 
$P_{{\rm e},h}$ is the 3D electron pressure profile in a single halo.
The term $W_{\gamma}(z, \ell)$ in Eq.~(\ref{eq:gamma_ell}) represents 
the kernel function of Eq.~(\ref{eq:gamma_kernel}) incorporated with the gamma-ray PSF effect.

\subsection{Information contents}

\begin{figure}
\begin{center}
       \includegraphics[clip, width=1.0\columnwidth]
       {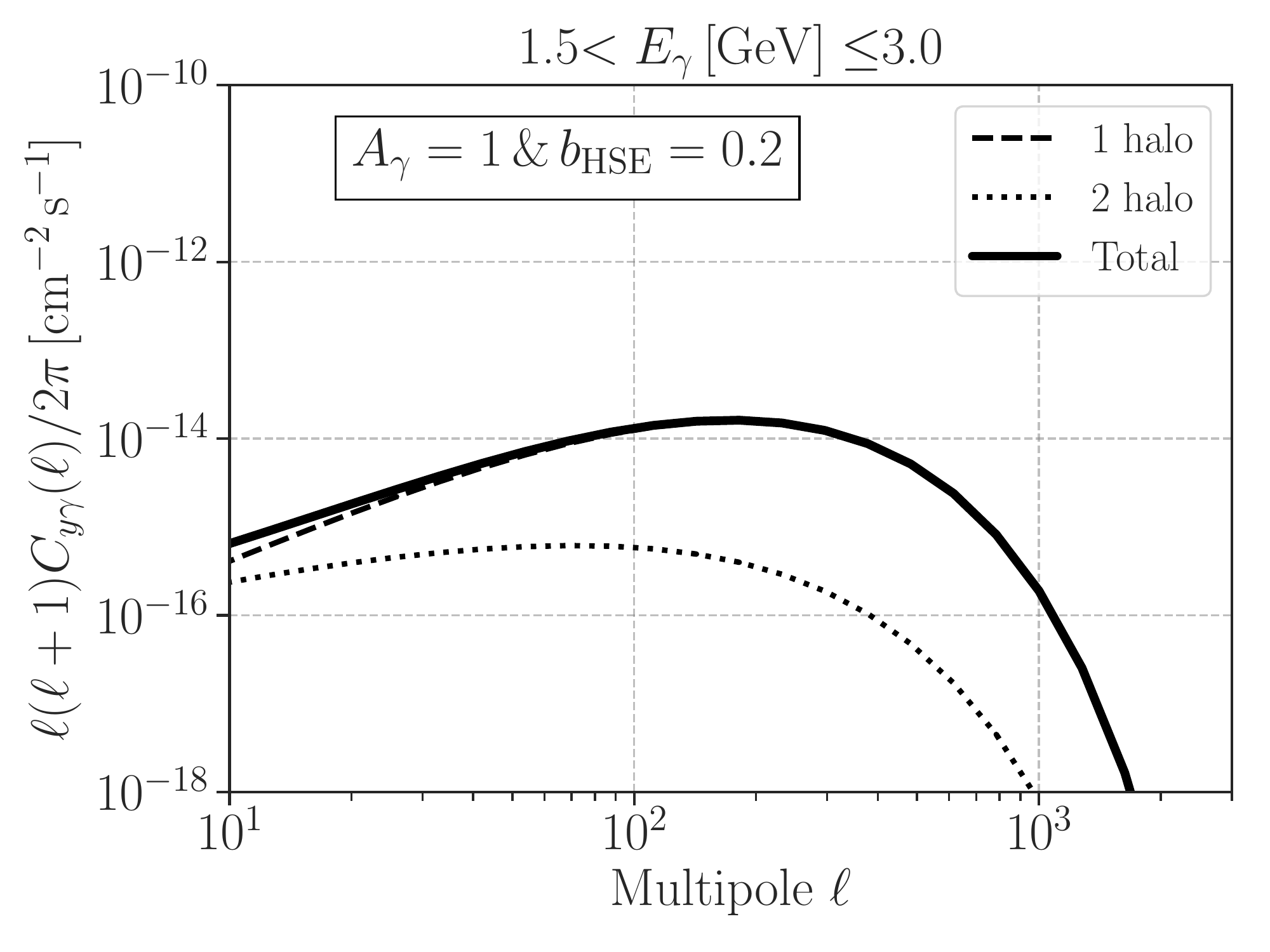}
     \caption{
     \label{fig:fiducial_Cl}
     Our fiducial model of the cross power spectrum between the UGRB intensity and the tSZ
     compton parameter. In this figure, we consider the gamma-ray energy range of $1.5 < E_{\gamma}\, [\rm GeV] \le 3.0$. 
     The dashed and dotted lines 
     show the one- and two-halo terms of the halo-model prediction, respectively.
     The dimensionless scale parameter in the gamma-ray emissivity in a single halo is set to 
     $A_{\gamma}=1$ in this figure. 
  }
    \end{center}
\end{figure}

We here summarize the information contents encoded in 
the cross power spectrum between the UGRB intensity and the tSZ Compton parameter. 
Figure~\ref{fig:fiducial_Cl} shows the analytic prediction 
of the cross power spectrum $C_{y\gamma}$ based on the halo-model approach. 
For this figure, we set the scale parameter in 
the gamma-ray intensity for single cluster-sized halos
(see Eq.~[\ref{eq:gamma_prof}]) to be $A_{\gamma}=1$ and assume the hydrostatic mass bias $b_{\rm HSE}=0.2$.
The dashed and dotted lines in the figure represent the one- and two-halo terms of 
the cross power spectrum, respectively. The clustering effect of neighboring halos on
$C_{y\gamma}$ would be important only at $\ell\simlt10$ and the two-point clustering
in single halos is more dominant over the wider range of multipoles.
This is because low-$z$ galaxy clusters would effectively contribute to the two-point clustering signal
and the angular size of the cluster becomes larger as the cluster redshift decreases.

\begin{figure}
\begin{center}
       \includegraphics[clip, width=1.0\columnwidth]
       {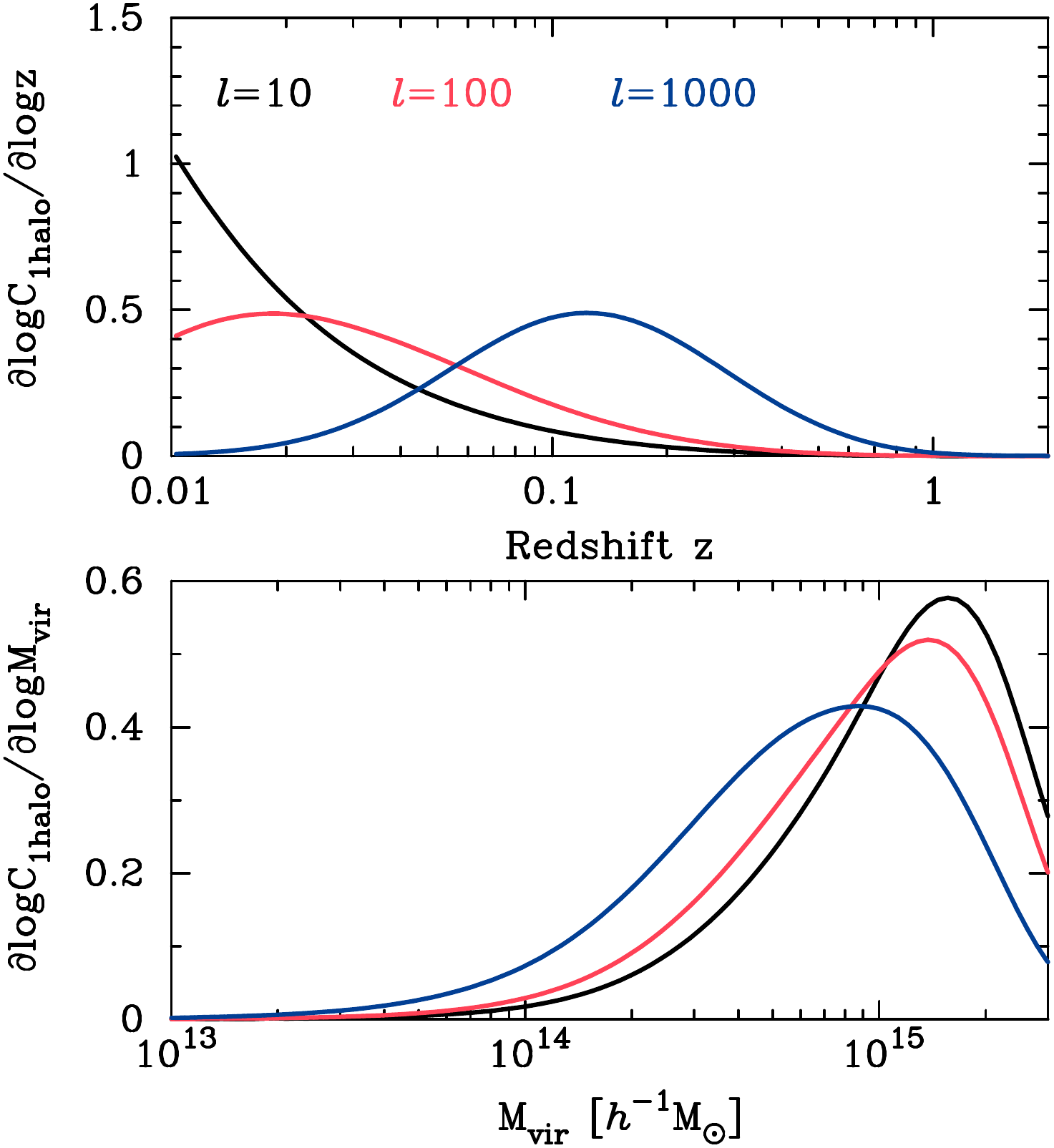}
     \caption{
     \label{fig:onehalo_each_z_M}
     The derivative of the one-halo cross power spectrum with respect to redshifts or halo masses.
     The upper panel shows the redshift dependence on the integrand of the one-halo term,
     while the bottom is for the mass dependence.
     In each panel, the black, red, and blue lines represent the results for three different multipoles
     $\ell=10, 100$, and 1000, respectively. 
  } 
    \end{center}
\end{figure}

To see effective redshifts and cluster masses probed by the cross power spectrum $C_{y\gamma}$,
we consider the derivative of the one-halo term to the redshift $z$ or the halo mass $M$:
For a given multipole $\ell$, these derivatives are given by
\beqa
\frac{\partial C^{\rm 1h}_{y\gamma}}{\partial z} &=& 
\frac{{\rm d}^2V}{{\rm d}z{\rm d}\Omega}\, \int_{M_{\rm min}}^{M_{\rm max}}\, {\rm d}M\, \frac{{\rm d}n}{{\rm d}M}\, |y_{\ell}(M,z) \gamma_{\ell}(M,z)|, \nonumber\\ \\
\frac{\partial C^{\rm 1h}_{y\gamma}}{\partial M} &=& 
\int_{z_{\rm min}}^{z_{\rm max}}\, \frac{{\rm d}^2V}{{\rm d}z{\rm d}\Omega}\, \frac{{\rm d}n}{{\rm d}M}\, |y_{\ell}(M,z) \gamma_{\ell}(M,z)|.
\eeqa
Figure~\ref{fig:onehalo_each_z_M} shows the derivatives for three different multipoles $\ell=10, 100$ 
and $1000$. The figure highlights that the cross power spectrum can contain the information 
of the galaxy clusters with their masses of $M\sim10^{15}\, h^{-1}M_{\odot}$ regardless of the multipoles.
At the degree-scale clustering (i.e. $\ell\simlt100$), the one-halo term is mostly determined 
by the contributions from the galaxy clusters at $z\simlt0.1$.
On the other hand, the cross correlation at smaller scales ($\ell\sim1000$) can 
probe the gamma-rays in galaxy clusters at $z\sim0.1$--0.2.
Since most gamma-ray studies of galaxy clusters concentrate on objects at $z\simlt0.1$ \cite{Ackermann:2013iaq, Zandanel:2013wea, Rephaeli:2015nca, Ackermann:2015fdi, Lisanti:2017qoz}, 
the cross-correlation analysis with the UGRB intensity and the tSZ Compton parameter is
a comprehensive approach to study gamma rays in the ICM at higher redshifts.

\begin{figure}
\begin{center}
       \includegraphics[clip, width=1.0\columnwidth]
       {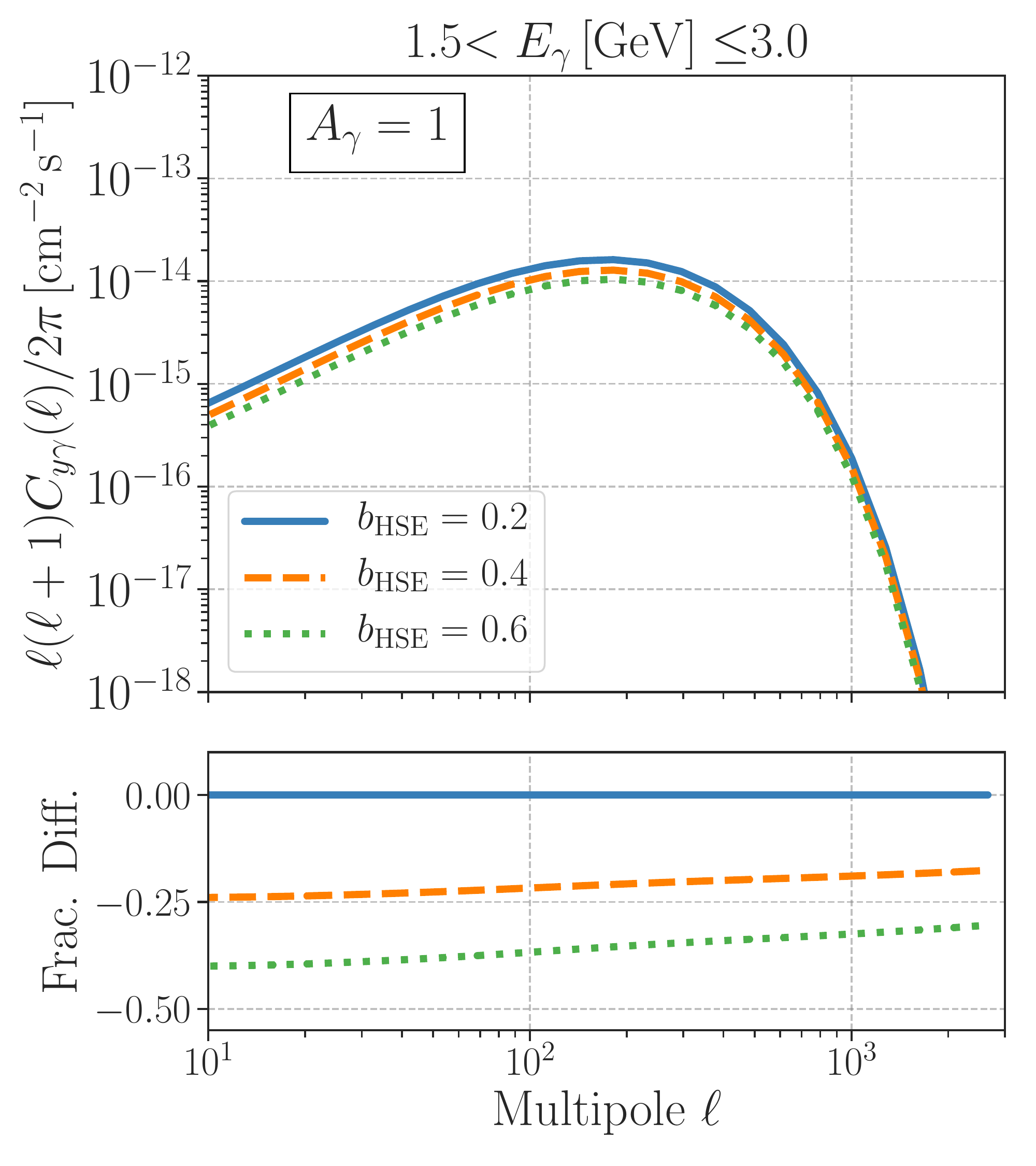}
     \caption{
     \label{fig:bHSE_dependence_Cl}
     The dependence of the cross power spectrum on the hydrostatic mass bias.
     The blue solid line shows the case of $b_{\rm HSE}=0.2$ as our baseline model,
     while the orange dashed and green dotted lines represents the results 
     with $b_{\rm HSE}=0.4$ and $b_{\rm HSE}=0.6$, respectively.
     The upper panel shows the cross power spectrum, while the bottom represents the fractional difference between the model with $b_{\rm HSE}=0.2$ and others.
  } 
    \end{center}
\end{figure}

The amplitude of $C_{y\gamma}$ should scale with $A_{\gamma}$.
Therefore, we can determine $A_{\gamma}$ with the measurement of the cross power spectrum
when assuming the cosmological model and the degree of the hydrostatic mass bias $b_{\rm HSE}$.
The exact value of $b_{\rm HSE}$ is still unclear even if we assume the concordance $\Lambda$CDM 
cosmology. Figure~\ref{fig:bHSE_dependence_Cl} shows the dependence on $b_{\rm HSE}$ of the cross power spectrum. We find that the shape of the power spectrum is almost unaffected by $b_{\rm HSE}$, but 
the amplitude shows a weak dependence of $b_{\rm HSE}$. 
Because a larger $b_{\rm HSE}$ leads to a smaller amplitude in the thermal pressure profile for a given halo mass $M$ [see Eq.~(\ref{eq:UPP_PLANCK})], the amplitude of the correlation is expected to decrease as $b_{\rm HSE}$ increases.
This indicates that the constraint of $A_{\gamma}$ by the cross power spectrum can depend on the assumed value of $b_{\rm HSE}$. In this paper, we consider a wide range of $b_{\rm HSE}$ from 0.1 to 0.7 
when constraining $A_{\gamma}$ with the measurement of the power spectrum (see Sec.~\ref{subsec:implication_cluster}).

It is worth noting that there should exist other contributions to the power spectrum from the clustering faint astrophysical sources at gamma-ray and microwave wavelengths.
In Appendix~\ref{apdx:blazar}, we examine the possible correlation between the main gamma-ray sources
and the tSZ effect by the ICM. We find that the contribution from the gamma-ray sources would be subdominant in the power spectrum, and thus, we ignore any possible 
cross-correlation signals arising from astrophysical sources.
Nevertheless, this treatment should provide a conservative upper limit on the parameter $A_{\gamma}$, since the correlation from the astrophysical sources is expected to be positive.

\section{\label{sec:data}DATA}

\subsection{\emph{Fermi}-LAT}\label{subsection:FermiDataSelection}

The data for this study were taken during the period August 4, 2008, to August 2, 2016, covering eight years. We used the current version of LAT data which is Pass 8\footnote{\url{https://fermi.gsfc.nasa.gov/ssc/data/analysis/documentation/Cicerone/Cicerone_Data/LAT_DP.html}} and the P8R3\_ULTRACLEANVETO\_V2 event class\footnote{The ULTRACLEANVETO event class comprises the LAT data with the lowest residual contamination that is publicly available.}. 
We also took advantage of a new event classification that divides the data into quartiles according to the localization quality of the events. In particular, we rejected the worst quartile denoted as \texttt{PSF0}. Furthermore, to reduce contamination from the Earth's albedo, we excluded photons detected with a zenith angle larger than 90$^{\circ}$. The data reduction procedure was done using version v11r5p3 of the Fermi Science Tools software package. Note that the selection cuts in our analysis are very similar to those introduced in Ref.~\cite{Ackermann:2018wlo}. The interested reader is referred to that article for validation tests and further checks on the data sample.

We analyzed LAT data in the energy range between 700 MeV and 1 TeV. The whole data set was subdivided into 100 logarithmically spaced ``micro" energy bins. For each micro-energy bin, we produced counts and exposure maps which were subsequently used to obtain raw flux maps. The resulting maps were spatially binned using the \texttt{Healpix}~\cite{Gorski:2004by} framework with $N_{\rm side}=512$. 
In this paper, we consider three energy bins of $0.7<E_{\gamma}\, [\rm GeV]\le1.5$, 
$1.5<E_{\gamma}\, [\rm GeV]\le3.0$, 
and $E_{\gamma} > 3 \, [\rm GeV]$ 
for cross-correlation analyses 
to study the gamma-ray energy dependence.
We also note that the effect of the energy dependent gamma-ray PSF is properly included in the theoretical model as in Ref.~\cite{Shirasaki:2014noa}, 
when we compare our model with the observed 
cross correlations.

\subsection{Compton-$y$ map by the \emph{Planck} satellite}\label{subsec:Planck_y}

To perform the cross-correlation analysis, 
we use the publicly available 
tSZ Compton map provided by the Planck collaboration. 
The Compton $y$ map has been constructed 
by the component separation of the Planck full mission data 
covering $30$ to $857 \, \mathrm{GHz}$ frequency channels \cite{Aghanim:2015eva}.
The original map is provided in \texttt{Healpix} format with $N_\mathrm{side} = 2048$,
but we degrade the map with $N_\mathrm{side}=512$ to be same as in the UGRB map.
Although 
the observed maps at multiple frequency bands have different beam properties,
we assume circularly symmetric Gaussian beam with the full-width half-mean (FWHM) beam size $\theta_\mathrm{FWHM} = 10.0 \, \mathrm{arcmin}$ for the Compton-$y$ map.
This Gaussian beaming effect is properly included in our theoretical model when we compare the model with the observed cross correlation.
For the map production, the Planck team examined two different component separation algorithm:
\texttt{MILCA} (Modified Internal Linear Combination Algorithm) \cite{2013A&A...558A.118H} and
\texttt{NILC} (Needlet Independent Linear Combination) \cite{2011MNRAS.410.2481R}.
Either is designed to find the linear combination of several components with optimal weight.
The weight is set so that the variance of the reconstructed map is minimized.
In this paper, we use the map constructed with \texttt{MILCA} as the fiducial map
because it has lower noise contribution at large scales.

\subsection{Masking}

When performing the cross-correlation analysis, we masked some regions to avoid any contamination
from resolved gamma-ray point sources and imperfect modeling of Galactic gamma-ray emission.
Namely, we masked all the extended and point-like sources listed in the 4FGL catalog~\cite{Fermi-LAT:2019yla}. For energies larger than 10 GeV  we also masked the 3FHL catalog~\cite{TheFermi-LAT:2017pvy} sources. The source mask takes into account both the energy dependence of the PSF and the brightness of each source. This is the same as in Ref.~\cite{Ackermann:2018wlo}, below we provide a brief description of the procedure proposed in that article.

For each micro-energy bin [$E\rm{_{i}}$, $E\rm{_{f}}$], we take the containment angle as given by $\rm{PSF}(E_{\rm{i}})$, which is in turn obtained as the mean of the three quartiles included in our data sample (PSF1, PSF2, PSF3). This value is subsequently used to define the radius of each source $r_{src}$. Conservatively, we take $r_{src}$ to vary from a minimum of 2$\times \rm{PSF}(E_{\rm{min}})$, for the faintest source with flux $F_{\rm{min}}$, to a maximum $F_{\rm{max}}$ of 5$\times \rm{PSF(E_{i})}$, for the brightest one. For sources with $F_{\rm{src}}$ somewhere in between these two extremes, we use a logarithmic scaling of the form~\cite{Ackermann:2018wlo}:
$$\frac{r_{\rm{src}}(F_{\rm{src}}, E_{\rm{i}}) - 2 \times \rm{PSF}(E_{\rm{i}})}{5 \times \rm{PSF}(E_{\rm{i}}) - 2 \times \rm{PSF}(E_{\rm{i}})} = \frac{\log(F_{\rm{src}}) - \log(F_{\rm{min}})}{\log(F_{\rm{max}})-\log(F_{\rm{min}})}$$
 As done in Ref.~\cite{Ackermann:2018wlo}, we also kept E$\rm{_{min}}$=8.3 GeV   for micro energy bins above 14.5 GeV.

We removed the Galactic diffuse emission (GDE) using the most up-to-date foreground emission model gll\_iem\_v07.fits. For this, we ran maximum likelihood fits in each micro-energy bin in which the flux normalization for the GDE model was free to vary. We also floated in the fits the normalization of an isotropic emission model (iso\_P8R3\_ULTRACLEANVETO\_V2\_v1.txt) accounting for the UGRB and possible cosmic-ray residuals in the data. Given that we are using the same amount of data used in the construction of the 4FGL catalog, it is well justified to fix all 4FGL point sources to their catalog values in the fitting procedure. The fits were performed with the \texttt{pylikelihood}\footnote{\url{https://fermi.gsfc.nasa.gov/ssc/data/analysis/documentation/Cicerone/}} routine within the Fermi Science Tools, which now provide support for likelihood analyses using maps in \texttt{Healpix} projection. In agreement with results in Ref.~\cite{Ackermann:2018wlo}, we found normalizations for the GDE that are within $1\sigma$ statistical uncertainty of the canonical values. Using our best-fit GDE model values, we constructed infinite-statistics model maps with the \texttt{gtmodel} tool in each energy bin. These were then subtracted from the raw flux maps. We applied the point source mask after this step to obtain the final UGRB maps.

As shown in Figure~\ref{fig:onehalo_each_z_M}, the ICM in low-$z$ galaxy clusters can affect the large-scale amplitude of the cross power spectrum.
To make our correlation analysis self-consistent, we apply circular masks around three nearby galaxy clusters at $z<0.01$. Those includes Virgo, Fornax, and Antlia clusters. We set the mask radius to be 8.0, 8.0 and 3.6
degree for Virgo, Fornax, and Antlia, respectively.
Note that these masks can cover the area beyond the virial region of individual nearby clusters \cite{Fouque:2001qc}.

\begin{figure}
\begin{center}
       \includegraphics[clip, width=1.0\columnwidth, viewport = 0 60 612 389]
       {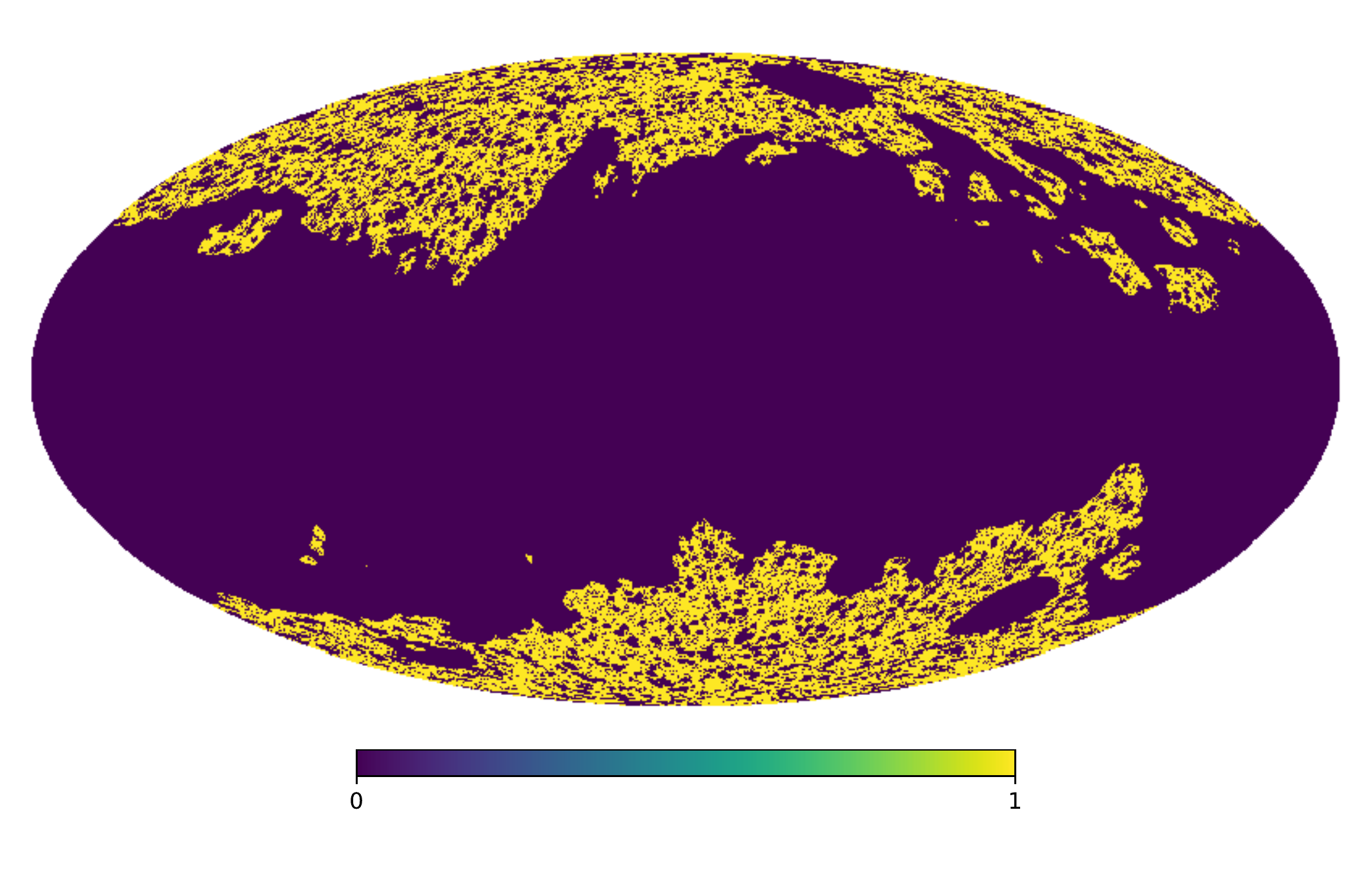}
     \caption{
     \label{fig:masks}
     Fiducial mask applied in our analysis for the $1.5<E_{\gamma}\, [\rm GeV]\le3.0$ energy bin. 
     Darker regions in this figure represent masks. 
     We mask the resolved gamma-ray and radio point sources and 
     the strong Galactic emission around the Galactic plane.
  } 
    \end{center}
\end{figure}

\begin{figure}
\begin{center}
       \includegraphics[clip, width=1.0\columnwidth]
       {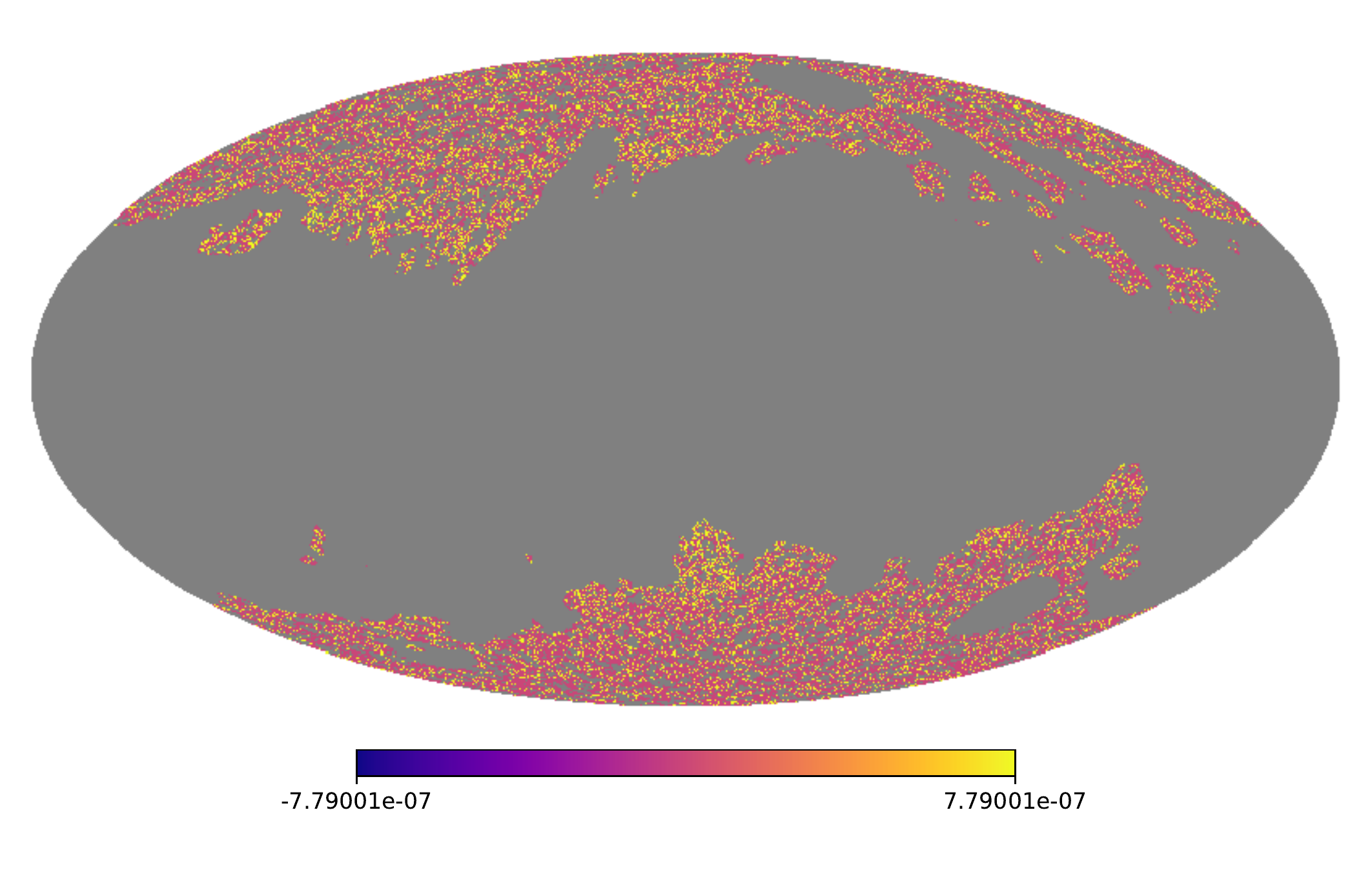}
       \includegraphics[clip, width=1.0\columnwidth]
       {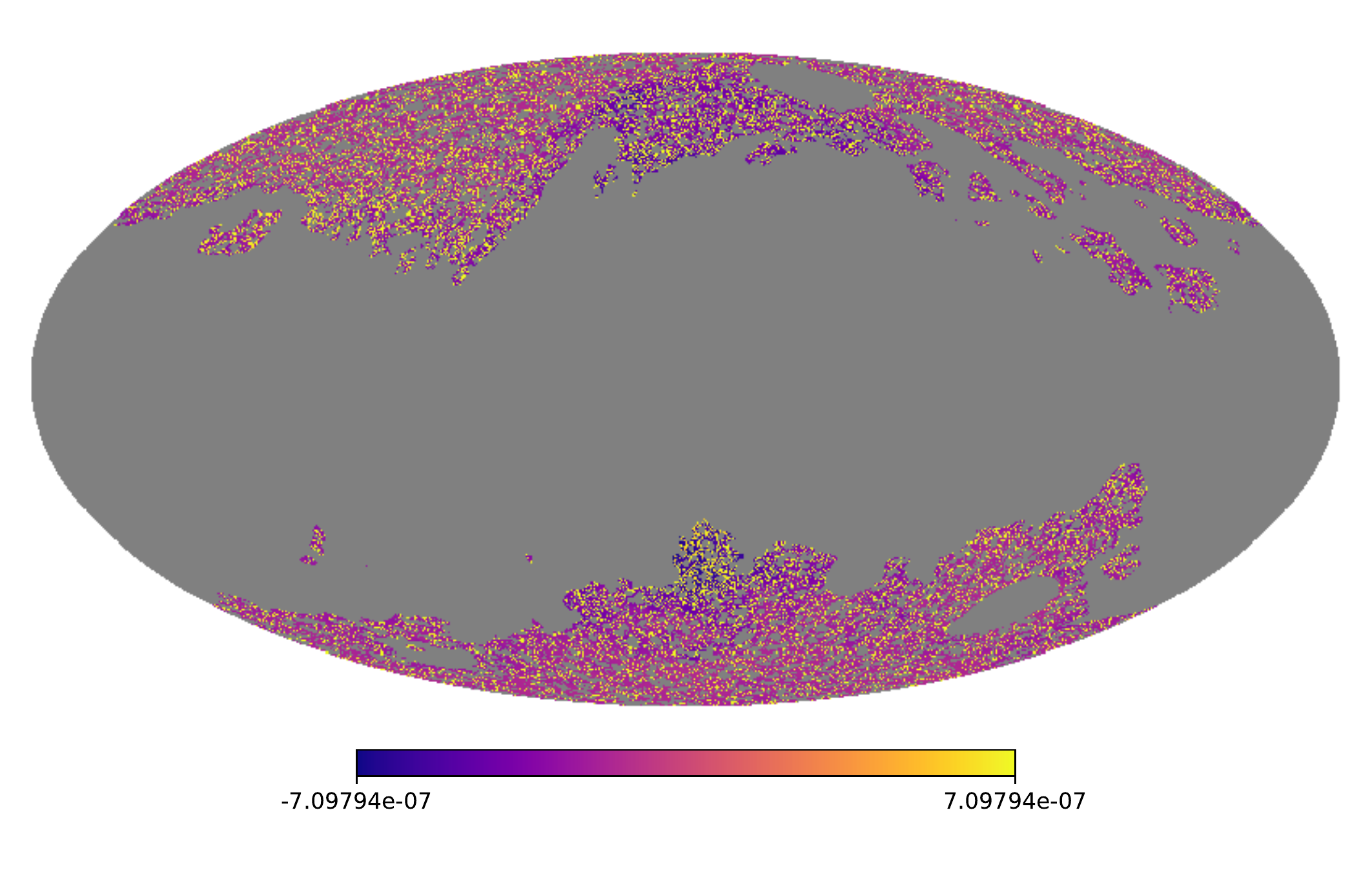}
       \includegraphics[clip, width=1.0\columnwidth]
       {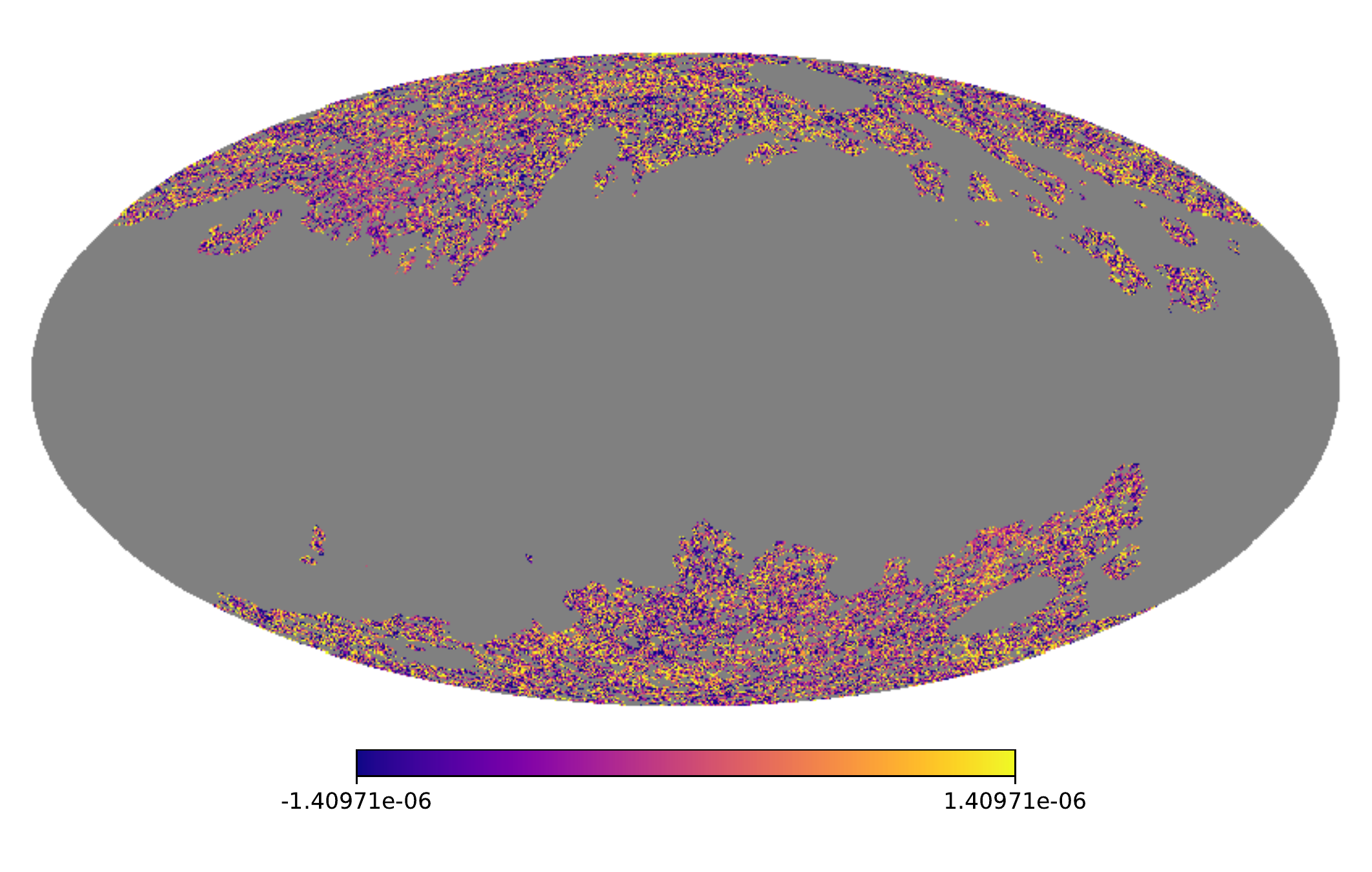}
     \caption{
     \label{fig:maps}
     Gamma-ray and compton $y$ maps in our analysis.
     The upper map shows the gamma-ray intensity map at $1.5<E_{\gamma}\, [\rm GeV]\le3.0$ before the subtraction of galactic components, while the middle map represents the UGRB counterpart. 
     The lower map is the {\tt MILCA} $y$ map provided by the Planck collaboration.
     Note that the gamma-ray intensity has units of ${\rm cm}^{-2}\, {\rm s^{-1}}$, while the $y$ map is dimensionless. In each map, the gray region shows the masked area.
  } 
    \end{center}
\end{figure}

For the microwave sky, we mask Galactic planes and point sources,
where strong radio emission component separation becomes unreliable.
We employ the 60\% Galactic mask and point source mask provided by Planck collaboration.
After placing the masks in the UGRB map at different gamma-ray energy bins, the sky coverage fraction of our data region is 11.1\%, 18.1\% and 22.1\% for the energy range of $0.7<E_{\gamma}\, [\rm GeV]\le1.5$, $1.5<E_{\gamma}\, [\rm GeV]\le3.0$, and $E_{\gamma}\, [\rm GeV]>3.0$, respectively.
Figure~\ref{fig:masks} shows our mask region used in the cross-correlation analysis for the gamma-ray energy bin of $1.5<E_{\gamma}\, [\rm GeV]\le3.0$, 
while Figure~\ref{fig:maps} shows the observed gamma-ray,
the UGRB and the Compton-$y$ maps.
It is worth noting that the CMB has a distinct component of diffuse Galactic emission called the Galactic ``Haze''. 
In practice, the Haze could still remain as a residual in the Planck $y$ map as well, and it would correlate with the Fermi Bubbles appeared in the middle in Figure~\ref{fig:maps} \cite{2013A&A...554A.139P}.
In Appendix~\ref{apdx:sys_err_powerspec}, we investigate the effect of the Fermi Bubbles and the Loop I structure on our power spectrum measurements by masking the known structures. 
There, we conclude that neither of the Fermi Bubbles nor the Loop I structure
compromises our results.

\subsection{Estimator of cross correlation}
\label{subsec:est_cross_corr}

We then estimate the cross power spectrum between the Fermi UGRB map and the Planck Compton $y$ map
using a pseudo-$C_{\ell}$ approach \cite{Hivon:2001jp}.
For this purpose, we make use of 
the publicly available tool {\tt PolSpice} \cite{Chon:2003gx, Szapudi:2000xj}.   
The algorithm properly deconvolves the power spectrum from mask effects in the maps of interest, 
but it is known not to be a minimum variance algorithm \cite{Efstathiou:2003dj}.  
In this sense, the associated covariance matrix is likely to overestimate the actual uncertainty, 
making the significance reported in this paper conservative.
We first measure the power spectrum in the multipole range from $\ell=10$ to 1000.
To mitigate possible mode-mixing effects caused by masks, 
we then average the measured power spectrum in 
10 logarithmic bins with a bin width of $\Delta \ln \ell = 0.46$.

The statistical uncertainty of the cross power spectrum $C_{y\gamma}$ can be
decomposed into two parts. One is the common Gaussian covariance term and it is given by
\beqa
{\rm Cov}_{\rm G}[C_{y\gamma_a}(\ell_{1}), C_{y\gamma_b}(\ell_{2})]
&=& \frac{\delta_{\ell_1 \ell_2}}{(2\ell_1+1)\Delta \ell \, \sqrt{f_{{\rm sky},a}f_{{\rm sky}, b}}} 
\nonumber\\&&\times
\Big[C_{yy}(\ell_1) C_{\gamma_a \gamma_b}(\ell_1) 
\nonumber \\ && + C_{y\gamma_a}(\ell_{1}) C_{y\gamma_b}(\ell_{1})\Big], \label{eq:cov_G}
\eeqa
where $C_{y\gamma_a}$ represents the cross power spectrum between the $y$ map and the $a$-th bin in the gamma-ray energy in the UGRB map, 
$C_{yy}$ is the auto power spectrum of the $y$ map,
$C_{\gamma_a \gamma_b}$ is the cross power spectrum between two different energy bins in the observed gamma-ray maps (including the Galactic emission),
and $f_{{\rm sky},a}$ is the sky fraction of 
the data region used in the cross-correlation analysis
at the $a$-th bin in the gamma-ray energy.
Note that each term in the right hand side in Eq.~(\ref{eq:cov_G}) is measurable with the {\tt PolSpice} algorithm.
Also, the {\tt PolSpice} calculates the
the right hand side in Eq.~(\ref{eq:cov_G}) including the Poisson noise.

The {\tt PolSpice} algorithm is not designed to provide the minimum-variance estimates, but the actual Gaussian covariance should be affected by the mode coupling due to sky masking \cite{Efstathiou:2003dj}. 
The public code of {\tt PolSpice} can provide the covariance matrix that take the geometric effects of mode coupling into account \cite{Challinor:2004pr} if the gamma-ray energy bins are identical, i.e., $a=b$ in Eq.~(\ref{eq:cov_G}).
Hence, we modify the Gaussian covariance term by using the covariance estimated by {\tt PolSpice} as \cite{Fornengo:2014cya}
\beqa
&& {\rm Cov}_{\rm G, mod}[C_{y\gamma_a}(\ell_{1}), C_{y\gamma_b}(\ell_{2})] \nonumber \\
&=&\left\{
\begin{array}{ll}
{\scriptstyle{\rm Cov}_{\rm P}[C_{y\gamma_a}(\ell_{1}), C_{y\gamma_b}(\ell_{2})]} & {\scriptstyle(a=b)} \\
{\scriptstyle \Gamma_{a}(\ell_1) \Gamma_{b}(\ell_2)
{\rm Cov}_{\rm G}[C_{y\gamma_a}(\ell_{1}), C_{y\gamma_b}(\ell_{2})]} & {\scriptstyle(a\ne b)} \\
\end{array} \right. ,\label{eq:cov_G_v2}
\eeqa
where ${\rm Cov}_{\rm P}$ is the covariance matrix provided by the {\tt PolSpice} code, and the correction factor $\Gamma_{a}(\ell_1)$ is defined as
\beqa
\Gamma_{a}(\ell) = \left(\frac{{\rm Cov}_{\rm P}[C_{y\gamma_a}(\ell), C_{y\gamma_a}(\ell)]}{{\rm Cov}_{\rm G}[C_{y\gamma_a}(\ell), C_{y\gamma_a}(\ell)]}\right)^{1/2}.
\eeqa
Note that Eq.~(\ref{eq:cov_G_v2}) includes the correlated scatters
among different $\ell$ bins.

Another contribution to the statistical error of $C_{y\gamma}$ is the four-point correlation function in the data region, referred to as the non-Gaussian covariance. We predict this non-Gaussian term based on the halo-model approach as in Sec~\ref{sec:model}. 
In the halo-model approach, the non-Gaussian covariance can be expressed as (e.g. see Ref.~\cite{2018MNRAS.480.3928M} for the cross-correlation between the Compton $y$ and galaxies)
\beqa
{\rm Cov}_{\rm NG}[C_{y\gamma_a}(\ell_{1}), C_{y\gamma_b}(\ell_{2})]
&=& \frac{1}{4\pi \sqrt{f_{{\rm sky},a}f_{{\rm sky},b}}} \int\, {\rm d}z\, \frac{{\rm d}^2V}{{\rm d}z{\rm d}\Omega}
\nonumber\\&&\times
\int\, {\rm d}M\, \frac{{\rm d}n}{{\rm d}M} \,  \nonumber \\ &&\times 
y_{\ell_1}\gamma_{a, \ell_1} y_{\ell_2}\gamma_{b, \ell_2}, \label{eq:cov_NG}
\eeqa
where $y_{\ell}$ and $\gamma_{\ell}$ are the Fourier transforms of the compton $y$ and the gamma-ray emissivity profiles for a single halo (see Sec~\ref{subsubsec:fourier_single_halo}).
Note that we omit the arguments of halo masses $M$ and redshifts $z$ for
$y_{\ell}$ and $\gamma_{\ell}$ in Eq.~(\ref{eq:cov_NG}) for simplicity.
In Appendix~\ref{apdx:non_G_err}, we show that the non-Gaussian error can be important for our measurements of the cross power spectrum at $\ell\sim100$. 

\section{\label{sec:res}RESULTS}

\subsection{Measurements of cross power spectrum}
\label{subsec:measurement}

\begin{figure}
\begin{center}
       \includegraphics[clip, width=0.9\columnwidth]
       {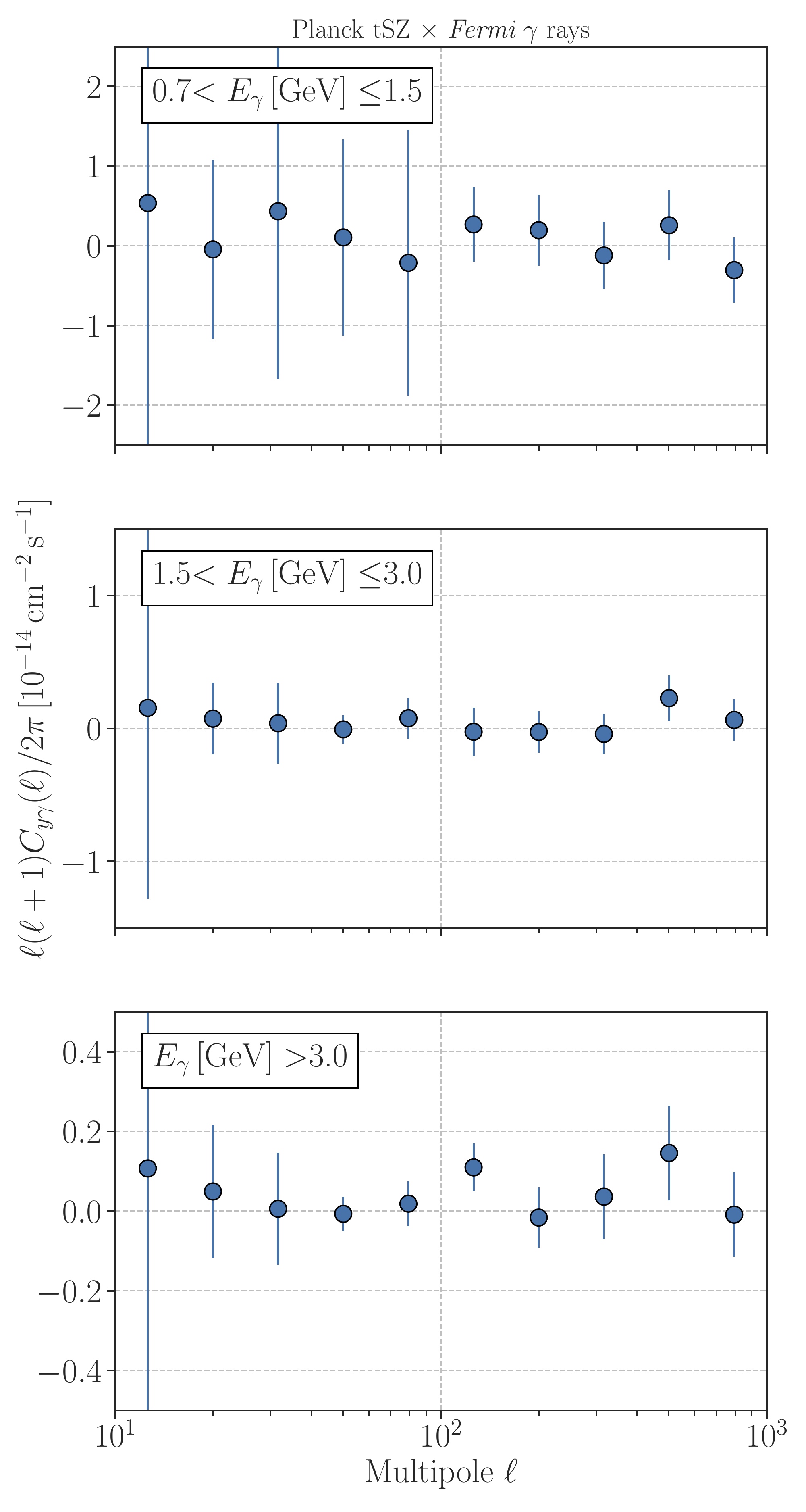}
     \caption{
     \label{fig:test_cl_diff_Emin}
     Measurement of cross power spectrum by varying the minimum gamma-ray energy in \emph{Fermi} UGRB map.
  } 
    \end{center}
\end{figure}

We summarize our measurement of the cross power spectrum between the Fermi UGRB and
the Planck Compton $y$ maps. 
Figure~\ref{fig:test_cl_diff_Emin} shows the measured power spectra 
for three different energy bins $0.7<E_{\gamma}\, [\rm GeV] \le 1.5$, $1.5<E_{\gamma}\, [\rm GeV] \le 3.0$ and $E_{\gamma}\, [\rm GeV]>3.0$.
The detection significance of the power spectra is commonly characterized as the signal-to-noise ratio,
which is defined by
\beqa
\left({\rm S}/{\rm N}\right)^2 &=& \sum_{a,b}\sum_{i,j}  
{\rm Cov}^{-1}_{\rm null}(\ell_{i}, \ell_{j}; a, b)\nonumber \\
&& \qquad \qquad \qquad \times
C_{y\gamma, a}(\ell_i)\, C_{y\gamma, b}(\ell_j), \label{eq:total_SN}
\eeqa
where $C_{y\gamma,a}(\ell_{i})$ is the cross power spectrum at the multipole $\ell_{i}$
for $a$-th energy bin in the UGRB map and ${\rm Cov}_{\rm null}$ is given by Eq.~(\ref{eq:cov_G_v2}) with $C_{y\gamma,a}=C_{y\gamma,b}=0$. Note that we set the non-Gaussian covariance to be zero in Eq.~(\ref{eq:total_SN}), because we define the significance testing a null detection.

\begin{table}
\begin{center}
\begin{tabular}{|c|c|}
\tableline
 & 
$10 < \ell < 1000$  \\ \hline
$0.7<E_{\gamma}\, [\rm GeV] \le 1.5$ 
& 2.59 (10) \\
$1.5<E_{\gamma}\, [\rm GeV] \le 3.0$
& 5.02 (10) \\
$E_{\gamma}\, [\rm GeV] > 3.0$
& 7.95 (10) \\ \hline
Combined 
& 10.76 (30) \\ \tableline
\end{tabular}
\caption{
\label{tab:total_SN} 
Summary of the significance of 
our cross-correlation measurements.
Second and third columns represent the $({\rm S}/{\rm N})^2$ defined in 
Eq.~(\ref{eq:total_SN}) and the numbers in brackets show the degree of freedom in the analysis. 
}
\end{center}
\end{table}

Table~\ref{tab:total_SN} represents the signal-to-noise ratio of our cross-correlation measurements.
We find that the power spectra at $\ell \simlt 100$ have larger
statistical uncertainties than at high-$\ell$s. This is
because the complex survey geometry induces mode coupling between different multipoles in a non-trivial manner.
Once taking into account the covariance between $\ell_{1}\ne \ell_{2}$,
we find that our measurement is consistent with a null detection.
In Appendix~\ref{apdx:sys_err_powerspec}, we examine three systematic effects in our measurement of the cross power spectrum 
to validate the null detection: imperfect modeling of Milky-way gamma-ray foregrounds, 
the inaccurate reconstruction of Compton $y$, and possible large-scale correlations between Galactic gamma rays with CMB maps.
In summary, we conclude that the cross power spectrum at $10<\ell<1000$ is minimally affected by these systematic uncertainties.

\subsection{Comparison with halo model}

\begin{figure}
\begin{center}
       \includegraphics[clip, width=1.0\columnwidth]
       {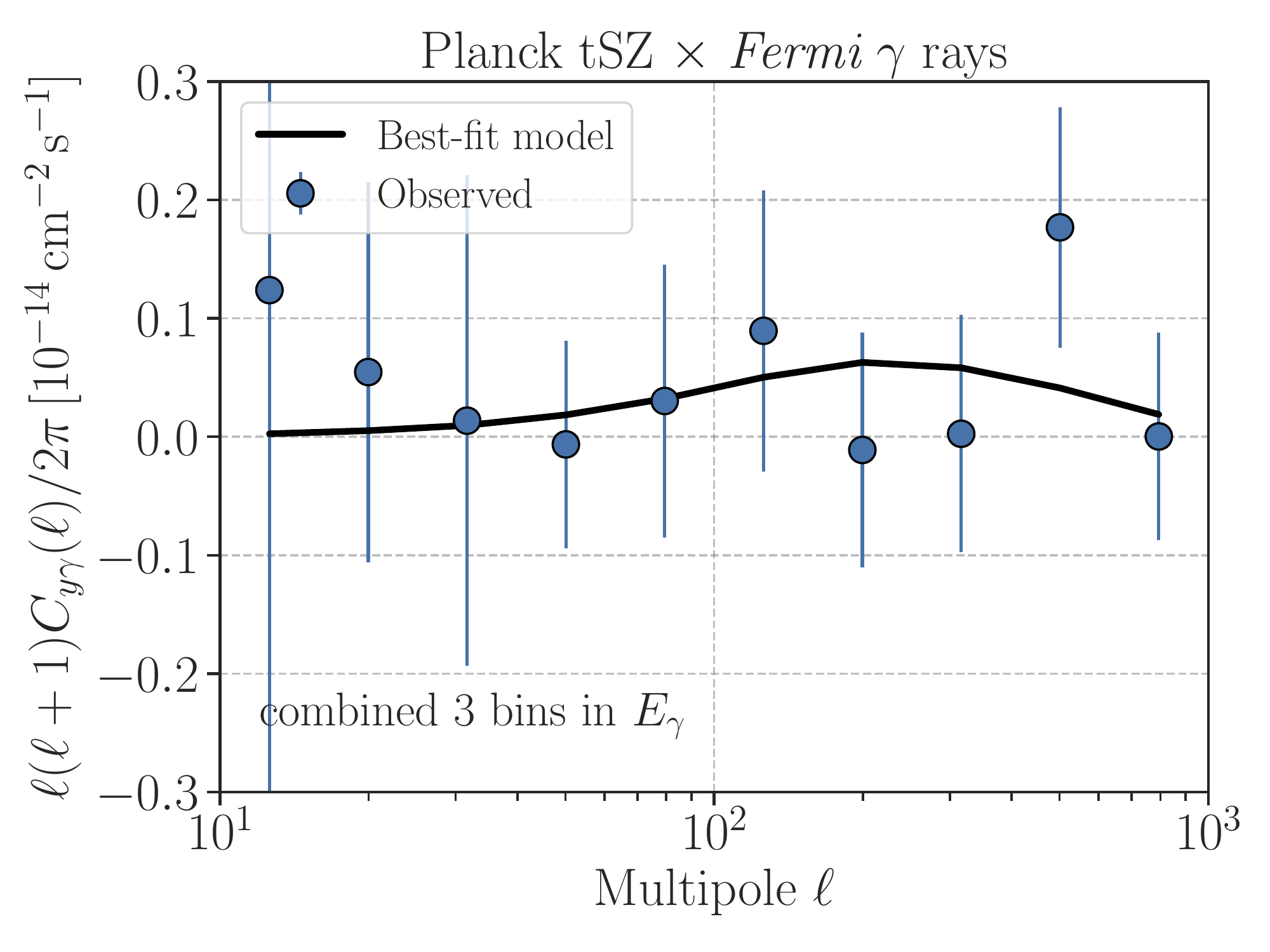}
     \caption{
     \label{fig:cl_vs_bfmodel_comb_E}
     Comparison of the observed cross power spectrum with our best-fit model. The gray hatched region is excluded in the likelihood analysis. In this figure, we combine the power spectra with 3 bins in the gamma-ray energy by using the minimum variance weight. See the text for the detail of the weight. Note that we assume the hydrostatic mass bias to be $20\%$ in this figure.
  } 
    \end{center}
\end{figure}

We compare our theoretical model of the UGRB-tSZ cross power spectrum with the measured signal. Since our halo-model prediction has two parameters $A_{\gamma}$ and $b_{\rm HSE}$,
we perform a likelihood analysis to find the best-fit model to the measurement. We infer the best-fit $A_{\gamma}$ to minimize the following log-likelihood for a given $b_{\rm HSE}$:
\beqa
-2 \log L &=& \sum_{a,b}\sum_{i,j}  
{\rm Cov}^{-1}_{\rm G+NG}(\ell_{i}, \ell_{j}; a, b;A_{\gamma})\nonumber \\ 
&& \qquad \times \left[C^{\rm obs}_{y\gamma_a}(\ell_{i})-C^{\rm mod}_{y\gamma_a}(\ell_{i}; A_{\gamma})\right] \nonumber \\
&& \qquad \times \left[C^{\rm obs}_{y\gamma_b}(\ell_{j})-C^{\rm mod}_{y\gamma_b}(\ell_{j}; A_{\gamma})\right], 
\label{eq:logL}
\eeqa
where ${\rm Cov}_{\rm G+NG}$ 
represents the covariance matrix defined by the sum of Eqs.~(\ref{eq:cov_G_v2}) and (\ref{eq:cov_NG}),
$C^{\rm obs}$ is the measured power spectrum, 
and $C^{\rm mod}$ is our model prediction.
In Eq.~(\ref{eq:logL}), the indices $a$ and $b$ run over the bins in the gamma-ray energy, 
while the indices $i$ and $j$ are for the bins in multipoles.
Note that the covariance matrix depends 
on the parameter $A_{\gamma}$ (see Eq.~\ref{eq:cov_NG}), 
but Ref.~\cite{2013A&A...551A..88C} points out that parameter estimates can be biased 
if one considers a parameter-dependence of covariance matrix in the Gaussian likelihood by including the term of
$\ln |\det {\rm Cov}|$ in Eq.~(\ref{eq:logL}).
To account for the parameter dependence of covariance in our likelihood analysis, 
we follow the same procedure as in Ref.~\cite{Makiya:2019lvm}. 
First, we infer the best-fit parameter by the likelihood analysis with 
covariance without the non-Gaussian term. Then, we compute the non-Gaussian covariance with the best-fit parameter and perform the likelihood analysis including the non-Gaussian covariance. We iterate this procedure until the best-fit parameter converges.
As the fiducial case, we assume $b_{\rm HSE}=0.2$ in this section.

\begin{figure}
\begin{center}
       \includegraphics[clip, width=1.0\columnwidth]
       {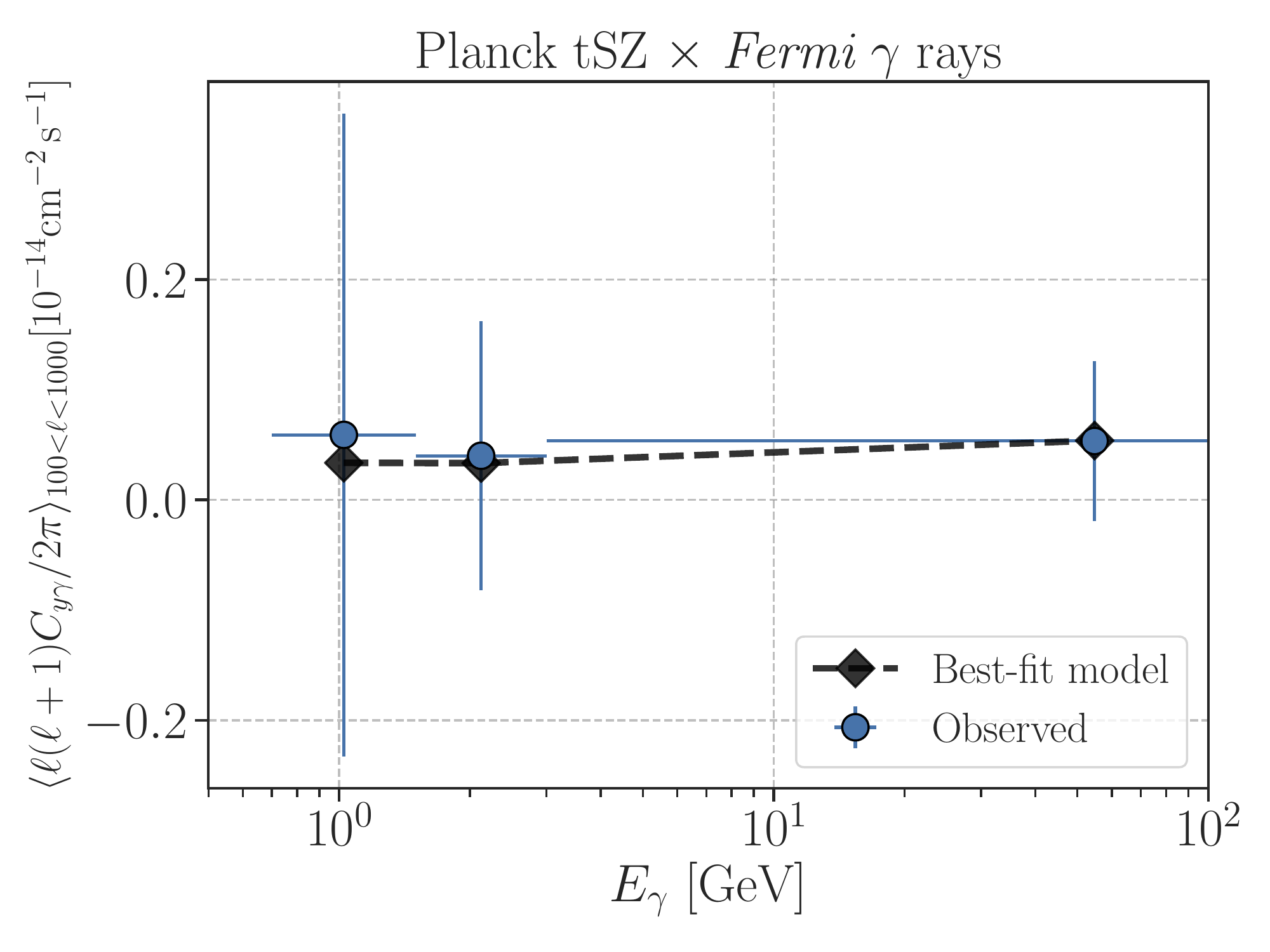}
     \caption{
     \label{fig:cl_vs_bfmodel_func_E}
     Similar to Fig.~\ref{fig:cl_vs_bfmodel_comb_E}, but we compare the cross power spectrum
     as a function of the gamma-ray energy bin.
  } 
    \end{center}
\end{figure}

Figure~\ref{fig:cl_vs_bfmodel_comb_E} shows the comparison 
with the measured power spectrum and the best-fit model.
In this figure, we combine the energy-dependent power spectra
by using the minimum variance weight (see Ref.~\cite{Shirasaki:2016kol} for a similar approach). 
The weight is then given by
\beqa
w_{a}(\ell) = \frac{1/{\rm Cov}_{\rm G+NG}(\ell, \ell, a, a)}{\sum_{b} 1/{\rm Cov}_{\rm G+NG}(\ell, \ell, b, b)},
\eeqa
and the weighted power spectrum is 
defined as $C^{\rm MV}_{y\gamma}(\ell) = \sum_{a} w_{a}(\ell)\, C_{y \gamma_a}(\ell)$.
We find the best-fit $A_{\gamma}$ to be $0.0348$
and our theoretical model can provide a reasonable fit 
to the observed power spectrum in the range of $10<\ell<1000$ as shown in the solid line in the figure.

Figure~\ref{fig:cl_vs_bfmodel_func_E} represents our fitting result as a function of the gamma-ray energy bin. 
For the visualization, we show the average power spectrum over the multipole range of $100<\ell<1000$
at each of gamma-ray energy bins.
The dashed line shows the best-fit model and it can explain the gamma-ray energy dependence of the measured power spectrum.

\subsection{Implications for galaxy clusters}
\label{subsec:implication_cluster}

\begin{figure}
\begin{center}
       \includegraphics[clip, width=1.0\columnwidth]
       {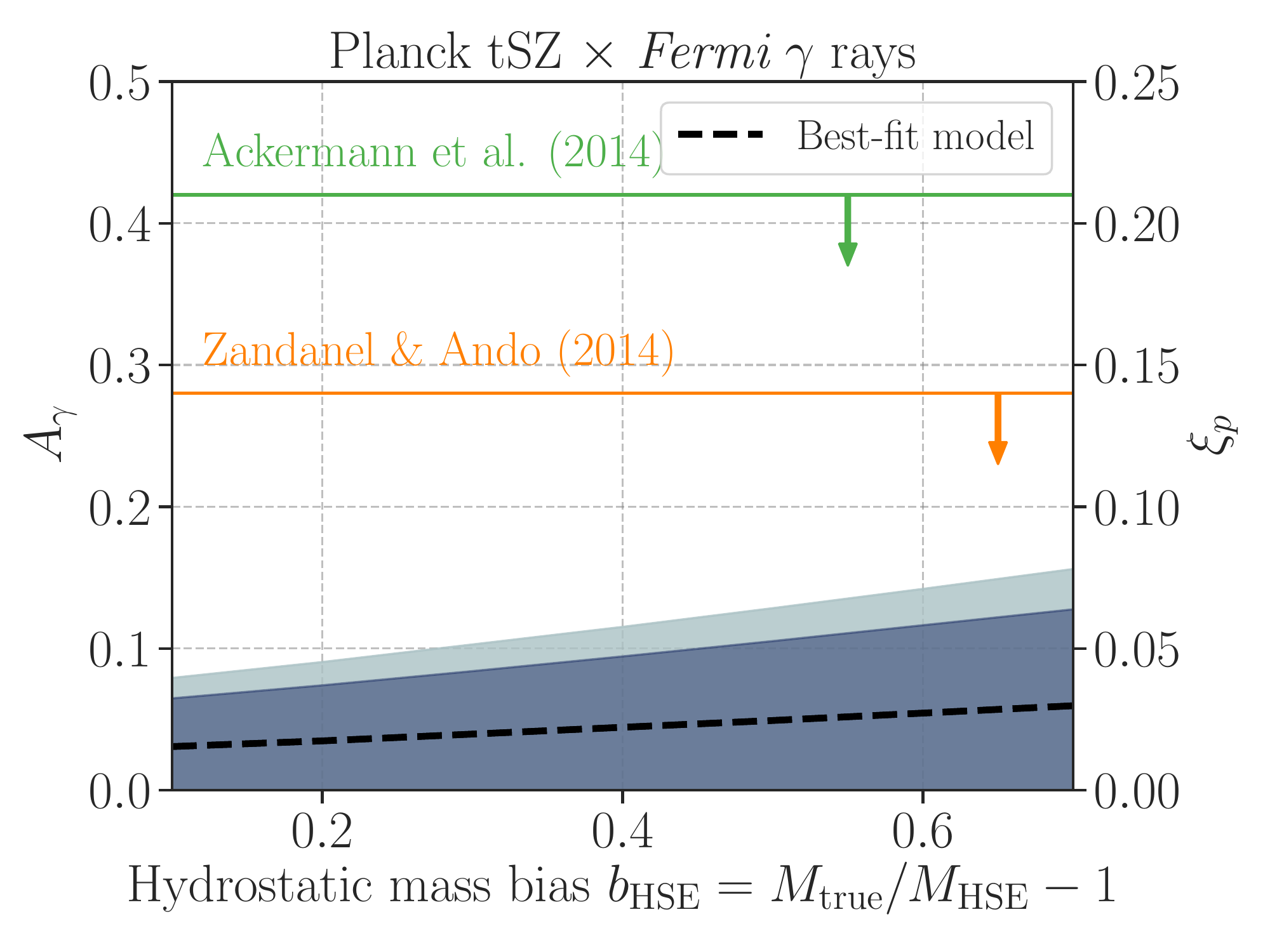}
     \caption{
     \label{fig:comparion_previous_studies}
     The comparison of our 
     constraints on the amplitude of cosmic-ray-included gamma-ray profile $A_{\gamma}$
     with respect to previous studies \cite{Ackermann:2013iaq, Zandanel:2013wea}.
     Our constraints are shown in the blue regions.
     The inner region (dark blue) shows the $1\sigma$ level, while 
     the outer one (dark grey) stands for the $2\sigma$ level.
    The right in the vertical axis shows the corresponding acceleration efficiency of cosmic ray protons at shocks $\xi_p$.
  } 
    \end{center}
\end{figure}

The comparisons 
between our model and the observed power spectrum
allows us to impose constraints on 
$A_{\gamma}$ for a given $b_{\rm HSE}$. 
Our likelihood analysis yields
the following 2$\sigma$-level constraints 
for three values of $b_{\rm HSE}$,
\beqa
A_{\gamma} &<& 0.0792 \, \, (b_{\rm HSE}=0.1), \\
A_{\gamma} &<& 0.0904 \, \, (b_{\rm HSE}=0.2), \\
A_{\gamma} &<& 0.102 \, \, (b_{\rm HSE}=0.3).
\eeqa
These constraints indicate that the acceleration efficiency of cosmic ray protons at shocks will be smaller than $\sim5\%$.
Figure~\ref{fig:comparion_previous_studies} summarizes the constraint 
on $A_{\gamma}$ as a function of $b_{\rm HSE}$
and compares our constraints with previous ones.
For the comparison with constraints obtained in previous works, 
we use Refs.~\cite{Ackermann:2013iaq} and \cite{Zandanel:2013wea}. 
The former performed a joint likelihood analysis searching for spatially extended gamma-ray emission 
at the locations of 50 galaxy clusters in four years of Fermi-LAT data, 
while the latter analyzed five-year Fermi-LAT data from the Coma galaxy cluster in the energy range 
between 100 MeV and 100 GeV.
Comparing 
against the constraints shown in these previous studies,
we find that our cross-correlation analysis can improve
the constraints on $A_{\gamma}$ 
by a factor of $\sim2-3$,
provided we assume the acceptable range of $b_{\rm HSE}$
in the Planck Compton-$y$ analyses \cite{Ade:2015fva, Bolliet:2017lha, 2018MNRAS.480.3928M}.

\begin{figure}
\begin{center}
       \includegraphics[clip, width=1.0\columnwidth]
       {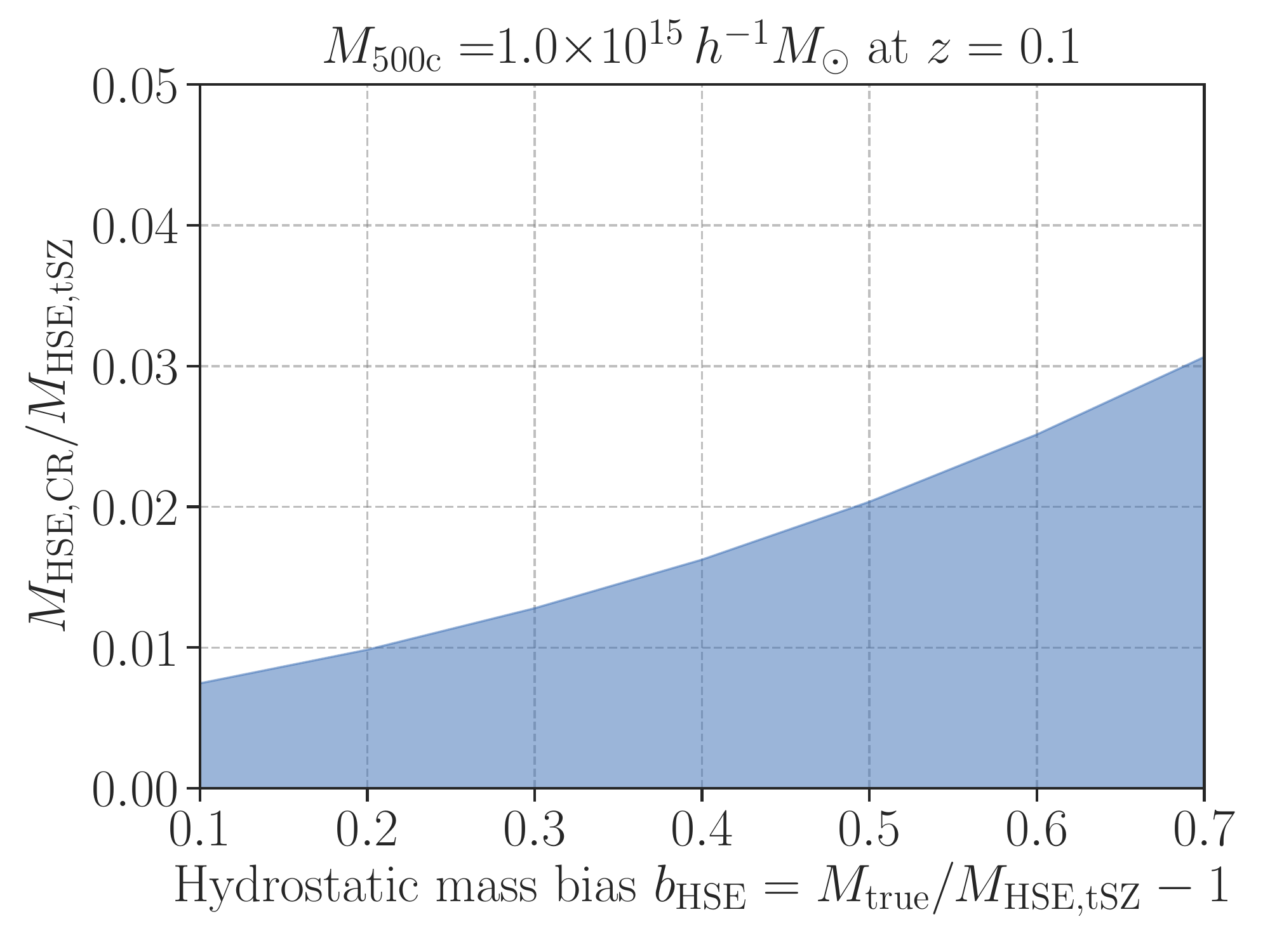}
     \caption{
     \label{fig:bias_CR_massive_cluster}
     The $2\sigma$-level upper limits on the cosmic-ray non-thermal pressure 
     as a function of hydro-static mass bias.
     To characterize the cosmic-ray-induced pressure in the model in Ref.~\cite{2010MNRAS.409..449P}, 
     we compute the hydrostatic mass defined by the cosmic-ray pressure in the unit of the thermal-pressure counterpart.
     In this figure, we assume a cluster with its mass of 
     $M_{\rm 500c}=10^{15}\, h^{-1}M_{\odot}$ at $z=0.1$.
     If the cosmic ray pressure is responsible to the observed hydrostatic mass bias $b_{\rm HSE}\sim0.3$, the quantity in the vertial axis should be close to $b_{\rm HSE}$.
  } 
    \end{center}
\end{figure}

The constraints 
on $A_{\gamma}$
in Figure~\ref{fig:comparion_previous_studies}
can convert the upper limit of 
the amount of non-thermal pressure induced by cosmic ray protons. For a given galaxy cluster with the mass $M$ at the redshift $z$, the cosmic-ray-induced pressure
can be expressed as $P_{\rm CR}(R) \propto A_{\gamma} C_{\gamma}(R)\, \rho_{\rm gas}(R)$ in the universal cosmic-ray model \cite{2010MNRAS.409..449P}, 
while the thermal electron pressure $P_{e}(R)$ 
is given by Eqs.~(\ref{eq:UPP_PLANCK}) and (\ref{eq:UPP}).
Thus, one can formally derive the hydrostatic mass 
using either $P_{\rm CR}$ or $P_{e}$.
Figure~\ref{fig:bias_CR_massive_cluster} 
shows the ratio of the hydrostatic mass defined 
by the cosmic-ray pressure and the thermal-pressure counterpart for the cluster mass 
$M_{\rm 500c}=10^{15}\, h^{-1}M_{\odot}$ at $z=0.1$. 
This figure 
shows that the cosmic-ray 
contribution to the cluster mass estimate 
should be smaller than 
the 1--3\% of the commonly-used hydrostatic mass by 
the thermal pressure for a wide range of $b_{\rm HSE}$.
This suggests that the cosmic-ray pressure can introduce
only a $\simlt1\%$ level of the mass bias if one adopts
the total hydrostatic mass bias to be $b_{\rm HSE}\sim0.3$.

\begin{figure}
\begin{center}
       \includegraphics[clip, width=1.0\columnwidth]
       {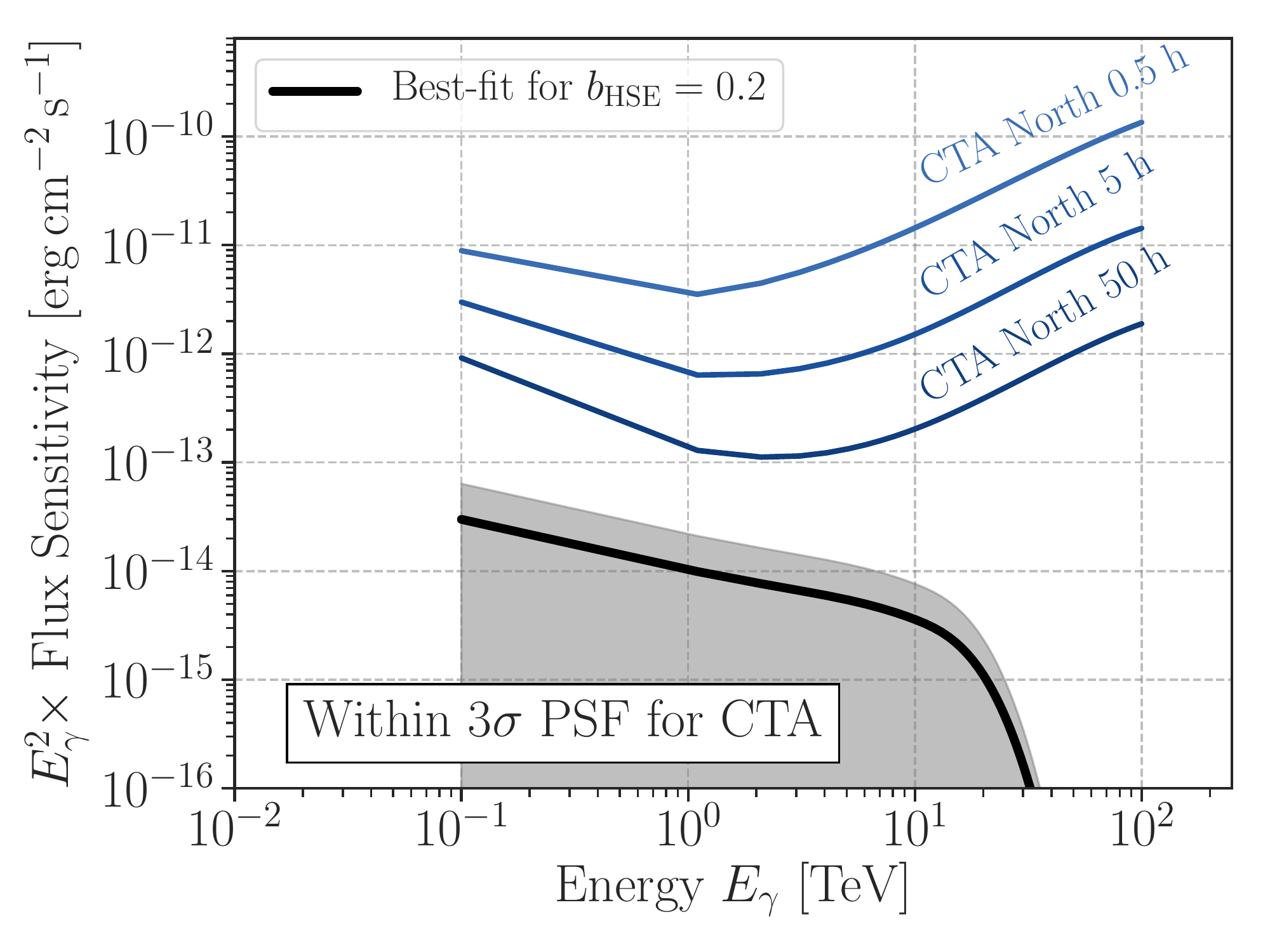}
     \caption{
     \label{fig:CTA_flux_Perseus}
     The expected gamma-ray flux 
     from the Perseus cluster
     by our model and comparison with the flux limit in the CTA experiment. In this figure, we consider the gamma-ray flux within the $3\sigma$-level PSF radius.
     For the model prediction, we set $A_{\gamma}=0.0348$ 
     assuming the hydrostatic mass bias $b_{\rm HSE}=0.2$. 
     The gray region represents the 
     $1\sigma$ statistical uncertainty inferred by our cross correlation analysis.
  } 
    \end{center}
\end{figure}

Finally, we study the detectability of 
the cosmic-ray-induced gamma rays from 
a nearby galaxy
cluster with the upcoming 
ground-based experiment by The Cherenkov Telescope Array (CTA)\footnote{\url{https://www.cta-observatory.org/}}.
As discussed in Ref.~\cite{Acharya:2017ttl}, the Perseus cluster is thought to be the best target for the detection of gamma rays by CTA. This is because the Perseus has a high ICM density at its center as well as it hosts the brightest radio mini-halo \cite{1990MNRAS.246..477P, Gitti:2002cd}.

The pion-decay-induced gamma-ray flux within the radius $R_{\theta}$ from a galaxy cluster is calculated by
\beqa
F(>E_{\gamma, {\rm min}}) &=& \frac{1}{D^2_{L}}
\int_{E_{\gamma, {\rm min}}}^{\infty} 
\frac{{\rm d}E_{\gamma}}{4\pi}\, \frac{{\cal S}(E^{\prime}_{\gamma},z)}{(1+z)^3}\,
e^{-\tau(E^{\prime}_\gamma, z(\chi))}
\nonumber \\
&& 
\qquad \times 
\int_{0}^{R_{\theta}}\,
2\pi R_{\perp} \, {\rm d}R_{\perp} \, 
\nonumber \\
&&
\qquad \times
\int_{-\infty}^{\infty} {\rm d}R_{\parallel}\, 
{\cal F}_{h}(R, M, z),
\eeqa
where $R=\sqrt{R^{2}_{\perp}+R^2_{\parallel}}$,
$D_{L}$ is the luminosity distance,
$E^{\prime}_{\gamma} = (1+z)E_{\gamma}$,
the energy spectrum ${\cal S}$ and 
the gamma-ray spatial distribution ${\cal F}_{h}$ are summarized in Section~\ref{subsec:ICM_single_halo}.
For the Perseus cluster, we assume its redshift to be 0.0183 and we adopt the model of the electron density 
constrained by the X-ray observation \cite{Churazov:2003hr}.
We also set the mass of the Perseus cluster 
by 1.2 times the hydrostatic mass obtained in Ref.~\cite{Churazov:2003hr} (i.e. we assume $b_{\rm HSE}=0.2$).
From the electron density $n_{e}$, 
we compute the gas density by 
$\rho_{\rm gas} = m_{p} n_{e} / (X_{H}X_{e})$.
To be conservative, we here ignore the gas clumpiness effect for the model prediction (i.e. $C_{\rm clump} = 1$).

Figure~\ref{fig:CTA_flux_Perseus} shows 
our model prediction of the gamma-ray flux from the Perseus cluster and the comparison with 
the expected flux limit by 
the CTA experiment\footnote{We infer the flux limit from the data in \url{https://www.cta-observatory.org/science/cta-performance/#1472563157332-1ef9e83d-426c}}.
The blue lines in the figure represent the flux limits as a function of the observational time, while 
the solid line is the prediction by our best-fit model.
According to a simple extrapolation, we expect 
that the flux limit with a 500-hour observation
will be comparable to the expected cosmic-ray induced gamma rays from the Perseus at 
$E_{\gamma, {\rm min}}\sim1\, {\rm TeV}$.
It would be worth noting that our model does not include the contribution from gamma-ray point sources in the Perseus cluster. To detect the ICM-induced gamma rays, one need to subtract the non-ICM contribution from real data as well. We leave investigations into more realistic gamma-ray analyses for future studies.

\subsection{Halo Model Uncertainties}
\label{subsec:model_uncertainty}

\begin{figure*}
\centering 
\includegraphics[clip,width=0.8\columnwidth]{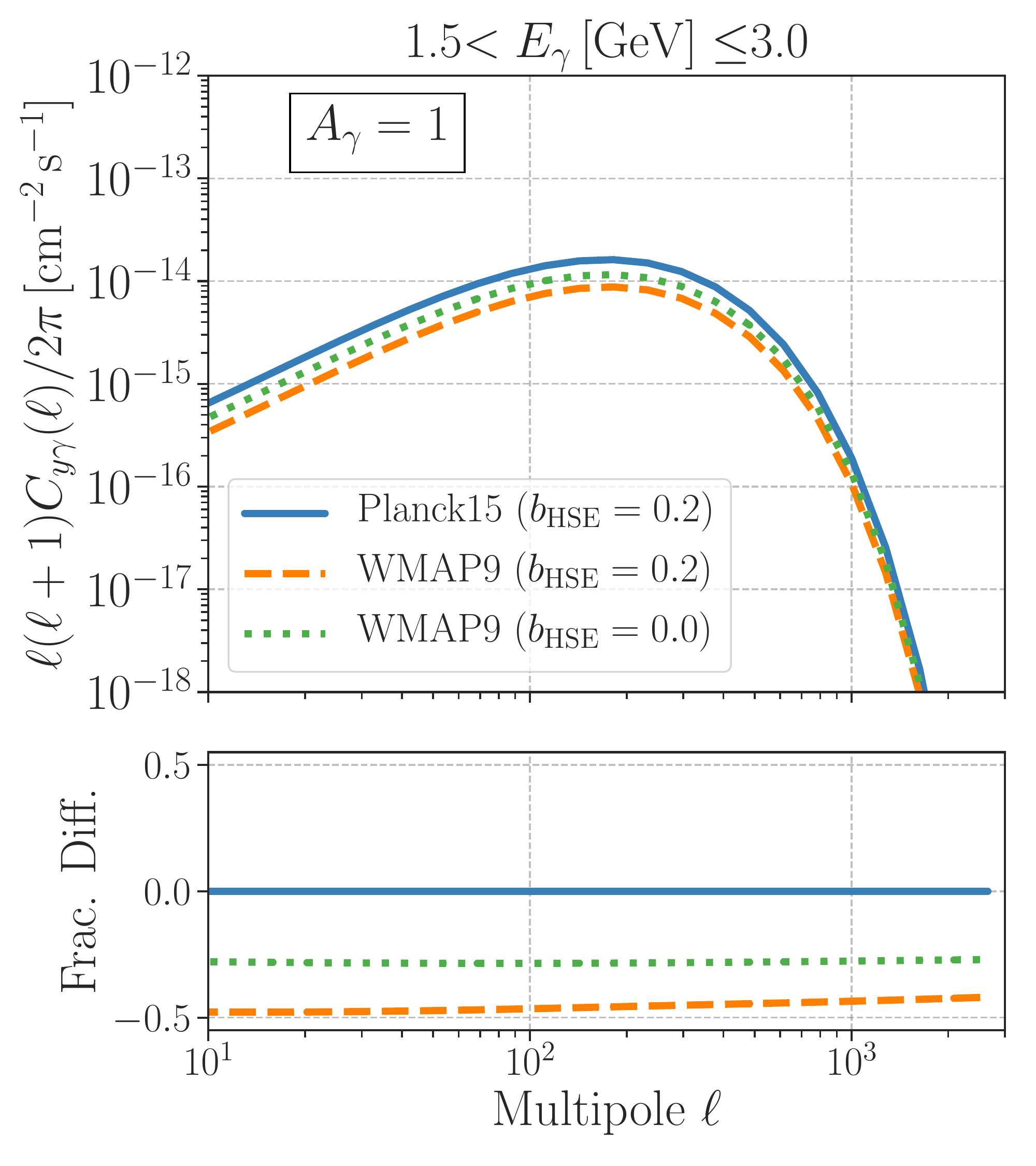}
\includegraphics[clip,width=0.8\columnwidth]{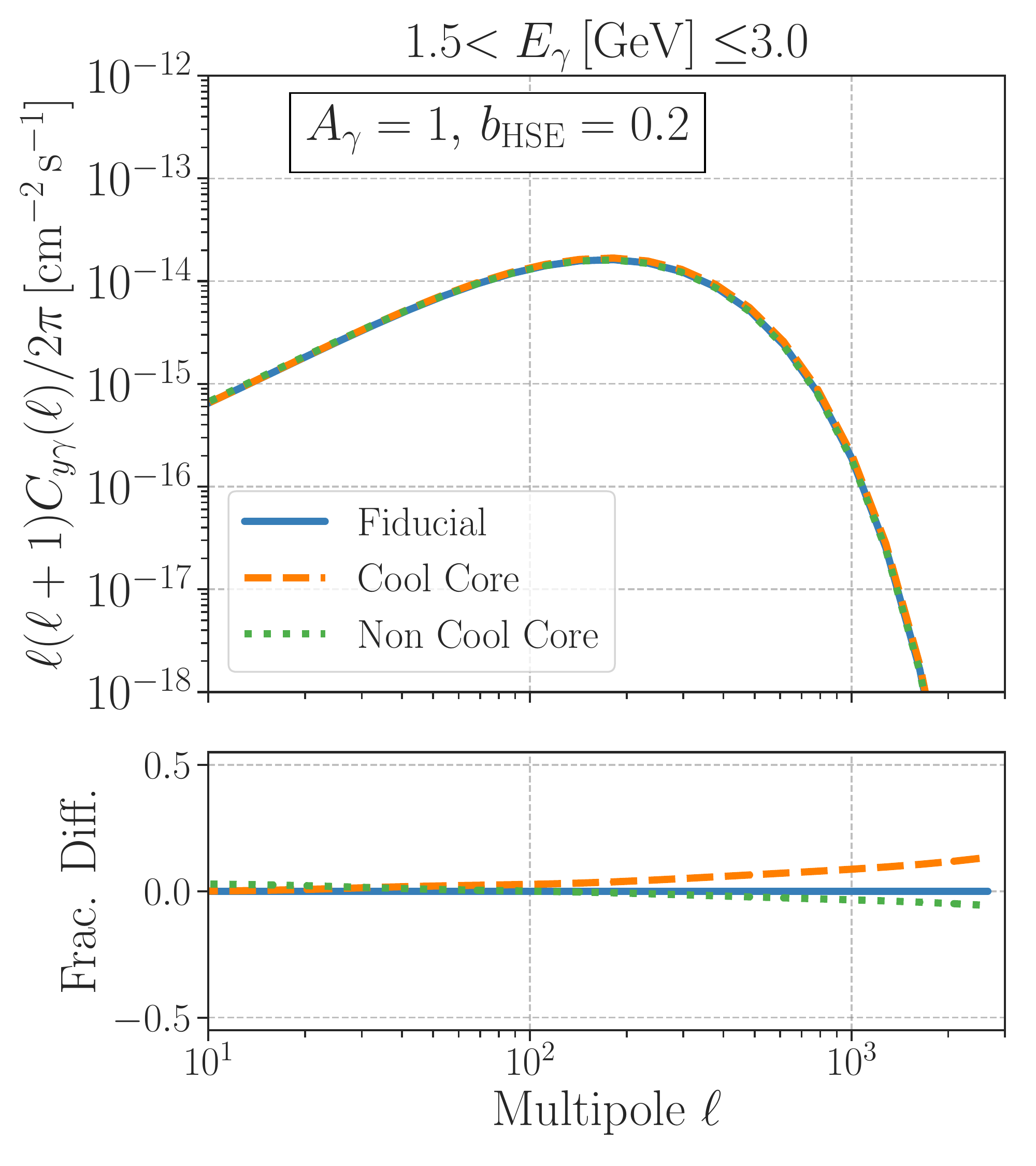}
\includegraphics[clip,width=0.8\columnwidth]{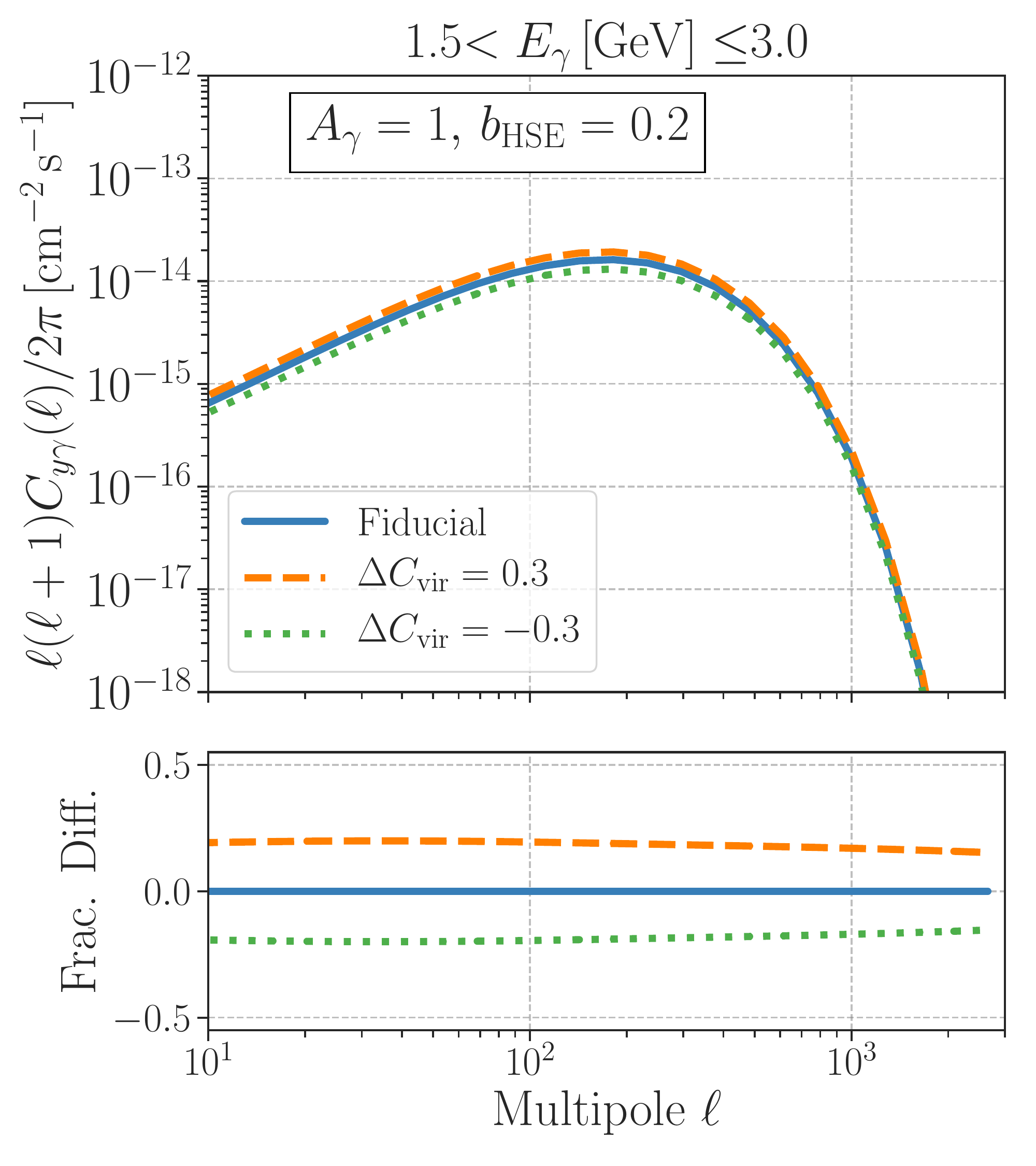}
\includegraphics[clip,width=0.8\columnwidth]{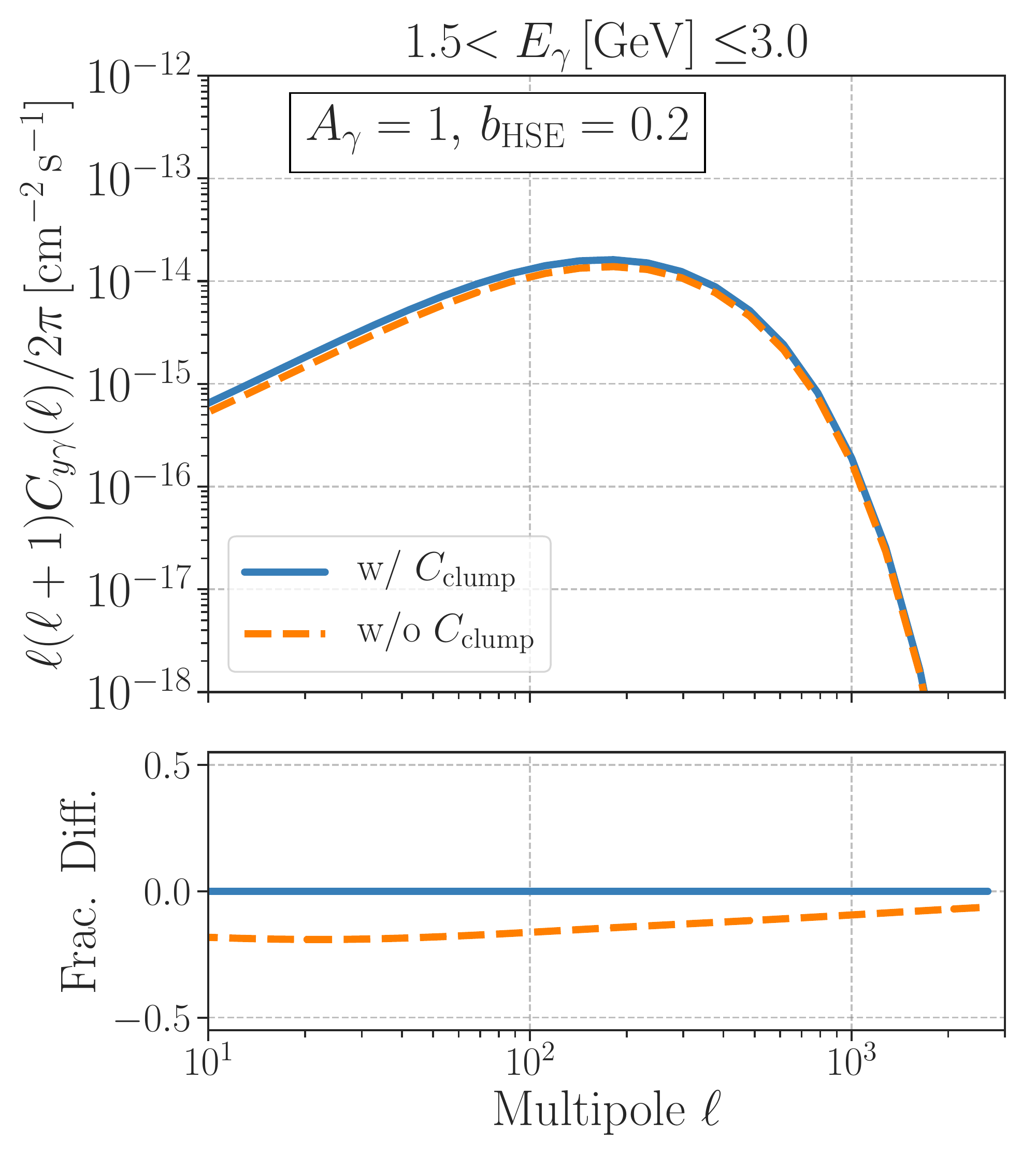}
\caption{
    Modeling uncertainties of the tSZ-UGRB cross power spectrum.
    In every panel, the upper portion shows power spectra in the gamma-ray energy range of 1.5-3.0 GeV with different model parameters,
    and the bottom potion shows the fractional difference to our fiducial model. {\it Top left}: The cosmological dependence. The blue line shows our fiducial model assuming the Planck15 cosmology and the hydrostatic mass bias $b_{\rm HSE}=0.2$, while the orange dashed line shows the model assuming the WMAP9 cosmology with $b_{\rm HSE}=0.2$. The green dotted line presents the model assuming the WMAP9 cosmology with $b_{\rm HSE}=0.0$, which is consistent with the recent constraint of cosmology and $b_{\rm HSE}$ inferred from the tSZ auto power spectrum in the Planck \cite{Bolliet:2017lha}.
    {\it Top right}: The dependence on the shape of ICM pressure profiles. The blue line shows our fiducial model, while the orange dashed and green dotted lines assume the pressure profiles for cool-core and non cool-core clusters, respectively. The parameters of the pressure profiles are taken from Ref.~\cite{2013A&A...550A.131P}. 
    {\it Bottom left}: The dependence on a parameter in gamma-ray profiles in PP10. We vary the amplitude of the mass-dependent term in the gamma-ray profiles (Eq.~\ref{eq:C_vir_gamma} or $C_{\rm vir}$) by $\pm30\%$. 
    {\it Bottom right}: The dependence on gas clumpiness. The blue line shows our fiducial model, while the orange dashed line represents the model without the effect of gas clumpiness. \label{fig:model_uncertainties}
	}
\end{figure*}

\rev{Our model based on the halo-model approach relies 
on several assumptions.
To assess the model uncertainties of the tSZ-URGB power spectrum, 
we consider four important elements in our model, and examine
the variations and uncertainties associated with them. Figure~\ref{fig:model_uncertainties} summarizes 
our findings.
In short, the cosmological parameters can cause a $\pm$30\%-level uncertainty, while the fitting function of the gamma-ray emission profile in Ref.~\cite{2010MNRAS.409..449P} and the gas clumpiness affect our modeling by $\pm$20\%. 
The detailed shape of the cluster pressure profile is found to be negligible for the current analysis. 
Hence, the total uncertainty in our model can amount
up to $30+20+20=70\%$. 
However, even considering the maximal $\pm$70\%-level uncertainty, we find that our constraint of $A_{\gamma}$ in Fig.~\ref{fig:comparion_previous_studies} is still tighter than previous limits.}

\subsubsection{Cosmological dependence}

\rev{The abundance of cluster-sized dark matter halos strongly depends on cosmological parameters \cite{2011ARA&A..49..409A}. 
Therefore, the assumed cosmology can affect our modeling 
of the tSZ-UGRB power spectrum. For our fiducial model, we adopt the cosmological parameters inferred from the CMB power spectra measured by Planck \cite{Ade:2015xua}. We refer this cosmological model as Planck15. For comparison, we also adopt the cosmological parameters constrained by the WMAP nine-year (WMAP9) data \cite{Hinshaw:2012aka}.
By assuming a reasonable value for hydrostatic mass bias $b_{\rm HSE}=0.2$, 
we find that the model based on the WMAP9 cosmology can differ from our fiducial model by a factor of 0.5. However, this comparison does not take into account another important constraint by the tSZ auto power spectrum.
The tSZ auto power spectrum can constrain the combination of cosmological parameters and $b_{\rm HSE}$ as $\sigma_8\, [\Omega_{\rm m0}/(1+b_{\rm HSE})]^{0.4}\, h^{-0.21}$ \cite{Bolliet:2017lha}.
To make the amplitude of the tSZ power spectrum consistent between Planck15 and WMAP9 models, we find that $b_{\rm HSE}=0.0$ is required for the WMAP9 model. When adding the prior information about cosmology and $b_{\rm HSE}$ expected from the tSZ power spectrum, we find that the WMAP9-based model is smaller than our fiducial model at a level of 30\%. Hence, we conclude that the current cosmological uncertainties can induce a $\pm$30\%-level uncertainty in our modeling of the tSZ-UGRB power spectrum.}

\subsubsection{Shape of cluster pressure profile}

\rev{It is known that the ICM pressure profile can depend 
not only on cluster mass but also other properties of 
individual clusters. 
The Planck observation of nearby galaxy clusters has found that
the pressure profile varies depending on whether a cluster has a central temperature drop\cite{2013A&A...550A.131P}. 
Clusters with central temperature drops 
are commonly called cool-core clusters. 
The parameters of the shape of pressure profiles for cool-core and non cool-core clusters have been constrained separately 
in Ref.~\cite{2013A&A...550A.131P}. 
We use those different parameters in modeling the tSZ-UGRB power spectrum and compare with our fiducial model. 
We find that the dependence of our modeling on the shape of pressure profile is small. It can induce at most a $\pm$10\%-level uncertainty at $\ell\simeq2000$. Since our likelihood analysis limits the multipole range to $10<\ell<1000$, we conclude that the modeling uncertainty associated with the pressure profile should be unimportant for the current analysis.}

\subsubsection{Fitting function of gamma-ray emission profiles}

\rev{Our model of the tSZ-UGRB power spectrum relies on the simulation results in Ref~\cite{2010MNRAS.409..449P}. The authors in Ref~\cite{2010MNRAS.409..449P} use a fitting formula 
for the gamma-ray emission profile as a function of 
cluster mass and radius.
Among the parameters in the fitting function, 
the amplitude of the mass-dependent term in the gamma-ray profile (Eq.~\ref{eq:C_vir_gamma} or $C_{\rm vir}$) 
appears to be subject to a $\pm$30\%-level uncertainty (see Figure~8 in Ref~\cite{2010MNRAS.409..449P}). We examine the impact of a $\pm$30\% difference in $C_{\rm vir}$ on the modeling of the tSZ-UGRB power spectrum. We find that the $\pm$30\%-level uncertainty in $C_{\rm vir}$ can change our prediction of the tSZ-UGRB power spectrum by $\pm$20\%.}

\subsubsection{Gas clumpiness}

\rev{The gas clumpiness effect (Eq.~\ref{eq:gas_clump}) can boost the expected cross power spectrum. We adopt the simulation-based model of $C_{\rm clump}$ as in Ref~\cite{Battaglia:2014cga}, while it has been poorly validated by actual observations. 
We examine the impact of gas clumpiness on our modeling and 
find that including the factor $C_{\rm clump}$ can increase the amplitude of the cross power spectrum by a factor of $\sim 20\%$.}

\section{\label{sec:con}CONCLUSION AND DISCUSSION}

We studied the gamma rays induced 
by the cosmic ray in the ICM 
using a cross-correlation analysis 
with the unresolved extragalactic gamma-ray background (UGRB)
and the thermal Sunyaev-Zel'dovich (tSZ) effect 
in the cosmic microwave background.
We developed a theoretical model of the cross-correlation signal based on the cosmic-ray model calibrated by the hydrodynamical simulation \cite{2010MNRAS.409..449P}.
We found that the cross power spectrum 
at the multipole $\ell\sim1000$ 
(or the equivalent angular scale being 
$\sim10\, {\rm arcmin}$) contains the information on
the cosmic-ray-induced gamma rays from the galaxy clusters 
outside the local Universe at $z=0.1$--0.2,
while clusters at $z<0.1$ are responsible for the signals at $\ell\sim 100$.

We also measured the cross power spectrum for the first time by 
using eight years of Fermi gamma-ray data and the publicly available tSZ map by Planck.
Our measurement is consistent with a null detection.
Comparing the observed power spectra with our theoretical model, we 
impose constraints on the acceleration efficiency of cosmic ray protons at shocks around the most massive objects in the Universe.
Our cross-correlation analysis sets the $2\sigma$-level upper limits of the acceleration efficiency to be $\sim7.8\%$.
This constraint is more stringent than previous ones
\cite{Ackermann:2013iaq, Zandanel:2013wea} by 
a factor of $\sim2-3$, while it is consistent with recent numerical studies \cite{Vazza:2016xuw, Ryu:2019uyk, Ha:2019ubf}.

Our constraint of the acceleration efficiency 
implies that the cosmic-ray pressure cannot be responsible for the observed hydrostatic mass bias in the tSZ-selected clusters \cite{Ade:2015fva}. 
We expect that the cosmic rays in the ICM will introduce a $\sim1\%$-level of the hydrostatic mass bias at most 
and it is smaller than the current limits of 
the hydrostatic mass bias
(e.g. see Refs~\cite{Ade:2015fva, 
Bolliet:2017lha, 2018MNRAS.480.3928M}).
Besides, we studied the future detectability of the pion-decay-induced gamma rays 
from the Perseus cluster with the upcoming CTA experiment.
Assuming the best-fit model to our cross-correlation measurement, we found a 500-hour observation with the CTA will be required to detect the gamma rays at the energy of $\sim1$ TeV from the Perseus. 

Our first measurement of the cross power spectra can be
further improved with the future ground-based CMB experiments \cite{Abazajian:2016yjj}, allowing to detect the cross power spectrum at $\ell\sim1000$ with a high significance level.
Such a precise measurement can reveal the nature of energetic components in the ICM as well as 
the physics of active Galactic nuclei 
(AGN) inside galaxy clusters.
Although our analysis ignores possible 
angular correlations caused by any astrophysical sources, 
it will become more important 
to understand the future precise measurement. 
A joint cross-correlation analysis among 
multi-wavelength data 
is one of the interesting approaches to constrain 
the nature of ICM as well as properties of any faint astronomical sources
(e.g. see Ref.~\cite{Shirasaki:2019ndb} for the ICM and Ref.~\cite{Shirasaki:2018wdq} for the astrophysical sources).
Future studies should focus on 
the development of accurate modeling 
of the ICM and astrophysical sources 
and optimal design of multi-wavelength data analysis.


\begin{acknowledgments}
This work was supported by 
MEXT KAKENHI Grant Numbers 18H04358 (M.S.), JP18H04340 and JP18H04578 (S.A.), and JSPS KAKENHI Grant Numbers 19K14767 (M.S.) and JP17H04836 (O.M. and S.A.).
Numerical computations were in part carried out
on Cray XC50 at Center for Computational Astrophysics,
National Astronomical Observatory of Japan. O.M. was also supported by World Premier International Research Center Initiative (WPI Initiative).
S.H.\ is supported by the U.S.\ Department of Energy under Award No.\ DE-SC0020262, NSF Grant No.\ AST-1908960, and NSF Grant No.\ PHY-1914409. 
\end{acknowledgments}

\appendix

\section{CROSS CORRELATION CAUSED BY GAMMA RAYS FROM ASTRONOMICAL OBJECTS}
\label{apdx:blazar}

\begin{figure}
\begin{center}
       \includegraphics[clip, width=1.0\columnwidth]
       {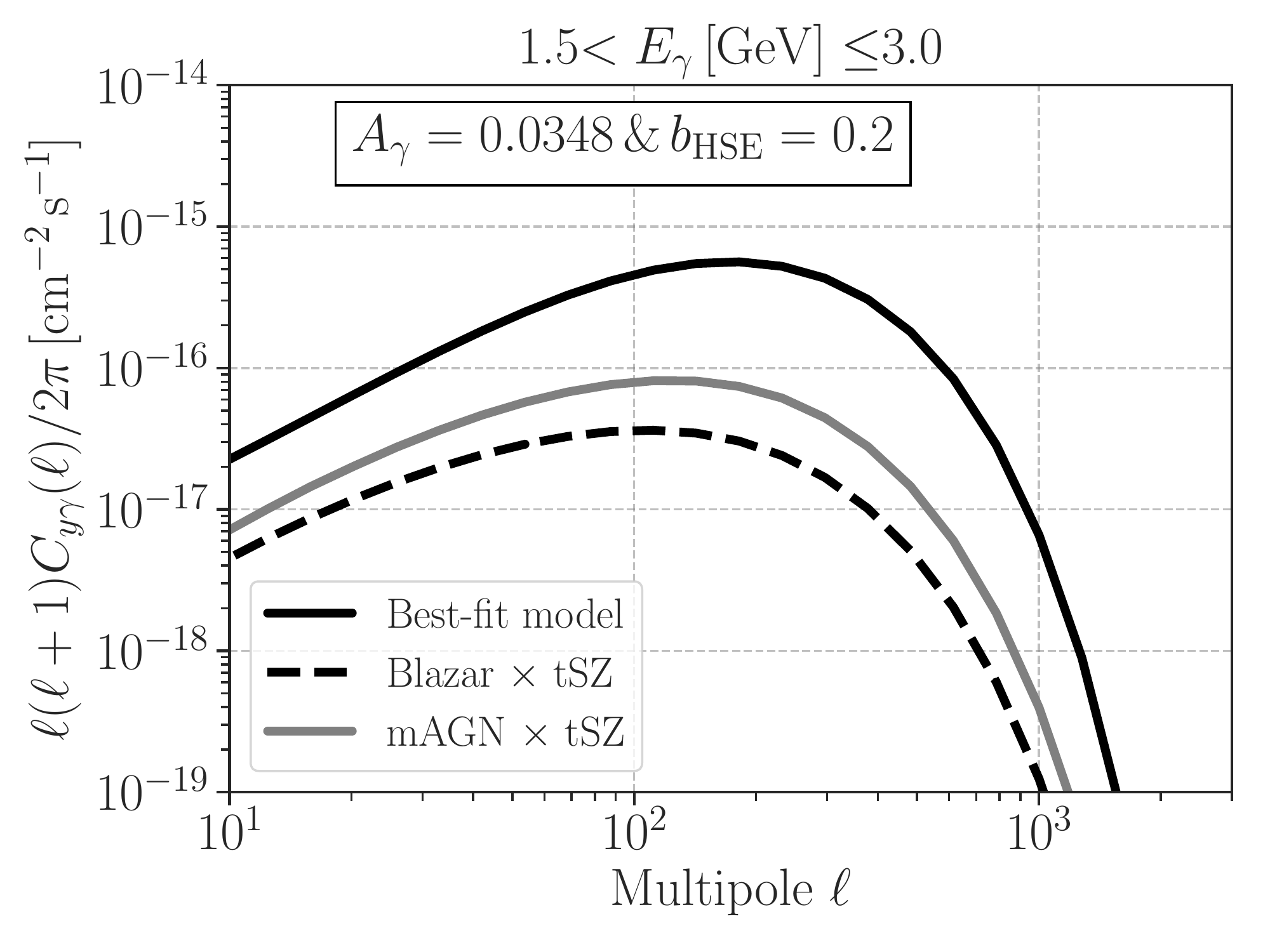}
     \caption{
     \label{fig:cl_Emin1.5GeV_blazar}
     The expected contribution to the cross power spectrum from the clustering of faint blazars and misaligned AGNs (mAGNs).
     The dashed line shows the signal caused by the blazars, 
     while the gray solid line represents the expected correlation with the mAGNs.
     For reference, the solid line shows the best-fit model of the cosmic-ray-induced signal to our power-spectrum measurement. In this figure, we included the beaming effect in the Planck compton $y$ map.
  } 
    \end{center}
\end{figure}

In the main text, we ignore possible correlations 
arising from the clustering of faint astrophysical sources
which cannot be resolved on an individual basis.
Among various astrophysical sources at gamma rays and microwave, 
blazars and misaligned AGNs (mAGNs) are expected to be potentially important in our analysis.
This is because faint blazar populations can be responsible for the UGRB at gamma-ray energies larger than
$\sim10$ GeV, while mAGNs can contribute significantly to the UGRB at $\sim1$ GeV \cite{Ajello:2015mfa}.
Also, blazars and mAGNs likely
reside in massive dark matter halos
(e.g., Refs~\cite{2014MNRAS.440.1527L, Allevato:2014qga, Shirasaki:2018dkz}).
The star forming activity in clusters can be a source of gamma rays in principle \cite{2012ApJ...755..117S}, however
we ignore this contribution in this paper. This is because galaxy clusters are known to have quenched star forming activity (e.g., see Ref.~\cite{2015ARA&A..53...51S}).

To evaluate the correlation between the gamma-ray emission from blazars and the tSZ effect by the ICM, 
we adopt the blazar model in Ref.~\cite{Shirasaki:2018dkz}.
In this model, the blazar is assumed to be a point source
and located at the center of a dark matter halo.
We also assume that each dark matter halo has at most one blazar.
The blazar gamma-ray luminosity function 
and the energy spectrum have been calibrated to 
the existing catalog of resolved gamma-ray blazars \cite{Ajello:2015mfa}.
We relate the gamma-ray luminosity of 
single blazars to their host halo mass by using a 
simple power-law model \cite{Camera:2014rja}.
The normalization and power-law index 
in the mass-luminosity relation have been determined so that 
the model can explain the abundance of X-ray selected AGNs \cite{Hutsi:2013hwa}.
We convert the gamma-ray luminosity to its X-ray counterpart
following Ref.~\cite{Inoue:2008pk}.

For mAGNs,
we adopt the model of Ref.~\cite{DiMauro:2013xta},
where the authors established a correlation
between the gamma-ray luminosity and 
the radio-core luminosity $L_{r,\rm core}$ at 5 GHz. 
Using the correlation together with the radio luminosity 
function of Ref.~\cite{Willott:2000dh}, we evaluate   
the gamma-ray luminosity function of mAGNs.
As for blazars, we assume that mAGN are point sources
residing in the center of dark matter halos, and that each 
dark matter halo can host at most a single mAGN.
We assume the mass-luminosity relation for mAGNs given in  Ref.~\cite{Camera:2014rja}.
To exclude blazars and mAGNs resolved by the Fermi telescope, 
we impose a flux cut at $E_{\gamma} > 100 \, {\rm MeV}$ of 
$2\times10^{-9}\, {\rm cm}^{-2}\, {\rm s}^{-1}$ in the model.
For details of our models for blazars and mAGNs, 
we refer the reader to Refs.~\cite{Camera:2014rja, Shirasaki:2018dkz, Hashimoto:2018ztv}.

Figure~\ref{fig:cl_Emin1.5GeV_blazar} shows the expected cross power spectrum between the gamma-ray emission from blazars and mAGNs and the tSZ effect by the ICM. In the figure, we consider gamma-ray data in 
the energy bin $1.5<E_{\gamma}\, [{\rm GeV}]\le3.0$. 
The solid line represents the best-fit model of the cross power spectrum by cosmic rays in the ICM to our measurement (see Section~\ref{subsec:measurement}), while 
the dashed and gray lines are for the contribution from blazars and mAGNs, respectively.
As seen in this figure, the contribution of 
the faint blazars and mAGNs to the UGRB-tSZ power spectrum is expected to be subdominant.
This is because the tSZ signal mostly 
comes from the most massive galaxy clusters (e.g., see Figure~\ref{fig:onehalo_each_z_M}), whereas faint astronomical objects would be mostly populated by smaller group-sized halos \cite{2014MNRAS.440.1527L, Allevato:2014qga}. 

\begin{figure}
\begin{center}
       \includegraphics[clip, width=1.0\columnwidth]
       {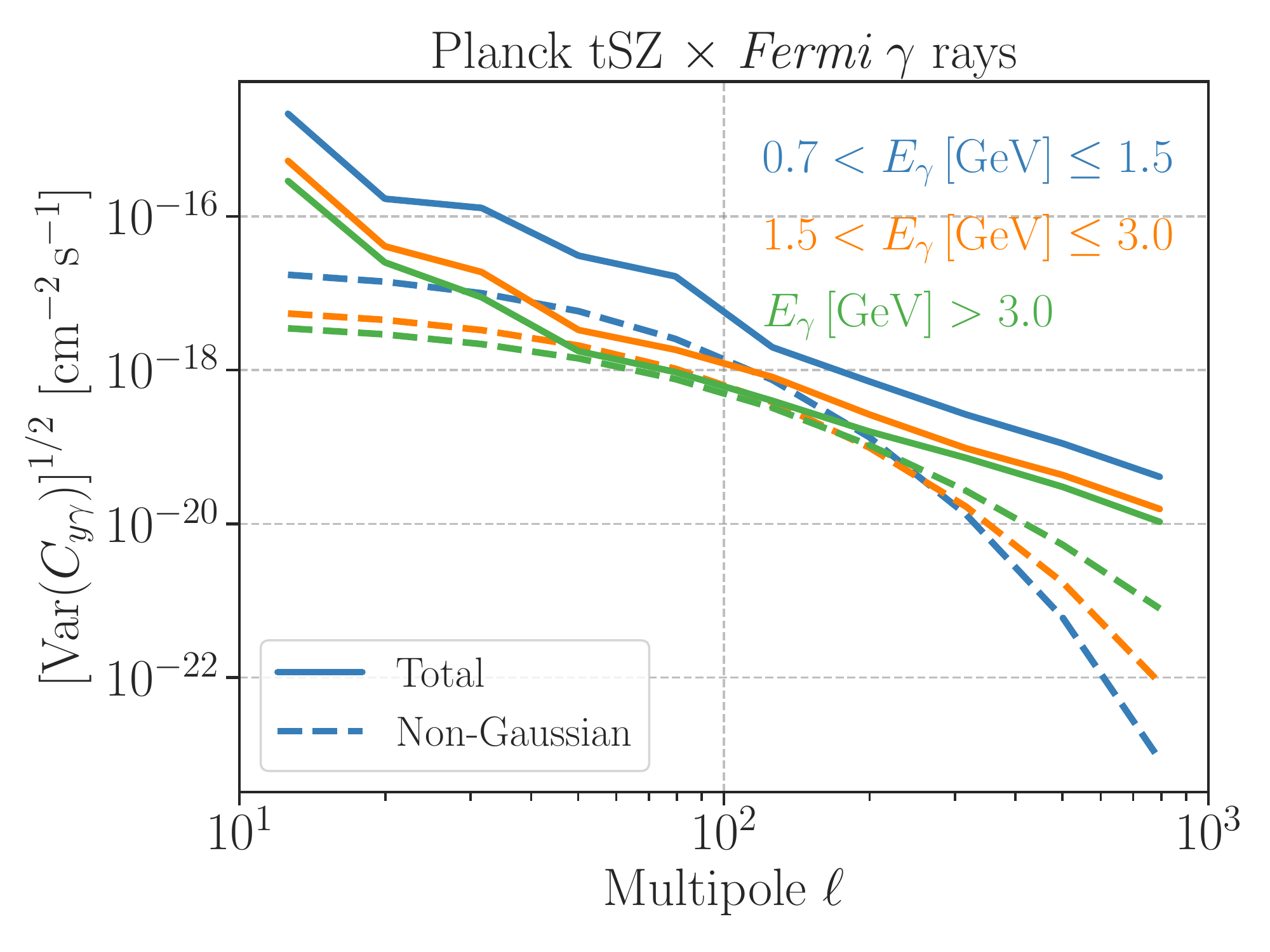}
     \caption{
     \label{fig:cl_sqvar}
     The statistical uncertainty in 
     our measurement of the cross power spectrum.
     The solid line shows the statistical error including
     the non-Gaussian contribution, while 
     the dashed one highlights the non-Gaussian error alone.
     The blue, orange, and green lines stand 
     for the analyses with 
     $0.7<E_{\gamma}\, [\rm GeV]\le1.5$, 
     $1.5<E_{\gamma}\, [\rm GeV]\le3.0$,
     and $E_{\gamma}\, [\rm GeV]>3.0$, respectively.
  } 
    \end{center}
\end{figure}

\section{STATISTICAL UNCERTAINTY 
OF UGRB-TSZ CROSS CORRELATION}
\label{apdx:non_G_err}

In this appendix, we show the effect of the non-Gaussian covariance in the UGRB-tSZ cross power spectrum, 
which is defined by Eq.~(\ref{eq:cov_NG}).
Figure~\ref{fig:cl_sqvar} shows the diagonal elements of the covariance matrix. The dashed line shows the non-Gaussian contribution arising from the four-point correlations in the data region. In this figure, 
we set $A_{\gamma}=0.0348$ and $b_{\rm HSE}=0.2$.
We find that the non-Gaussian error is subdominant in the diagonal elements of the covariance in the range of $\ell\simgt100$,
while it can become comparable to the conventional Gaussian error at $\ell\sim100$.

\section{SYSTEMATIC UNCERTAINTY OF UGRB-TSZ CROSS CORRELATION}
\label{apdx:sys_err_powerspec}

\begin{figure*}
\begin{center}
       \includegraphics[clip, width=1.0\columnwidth]
       {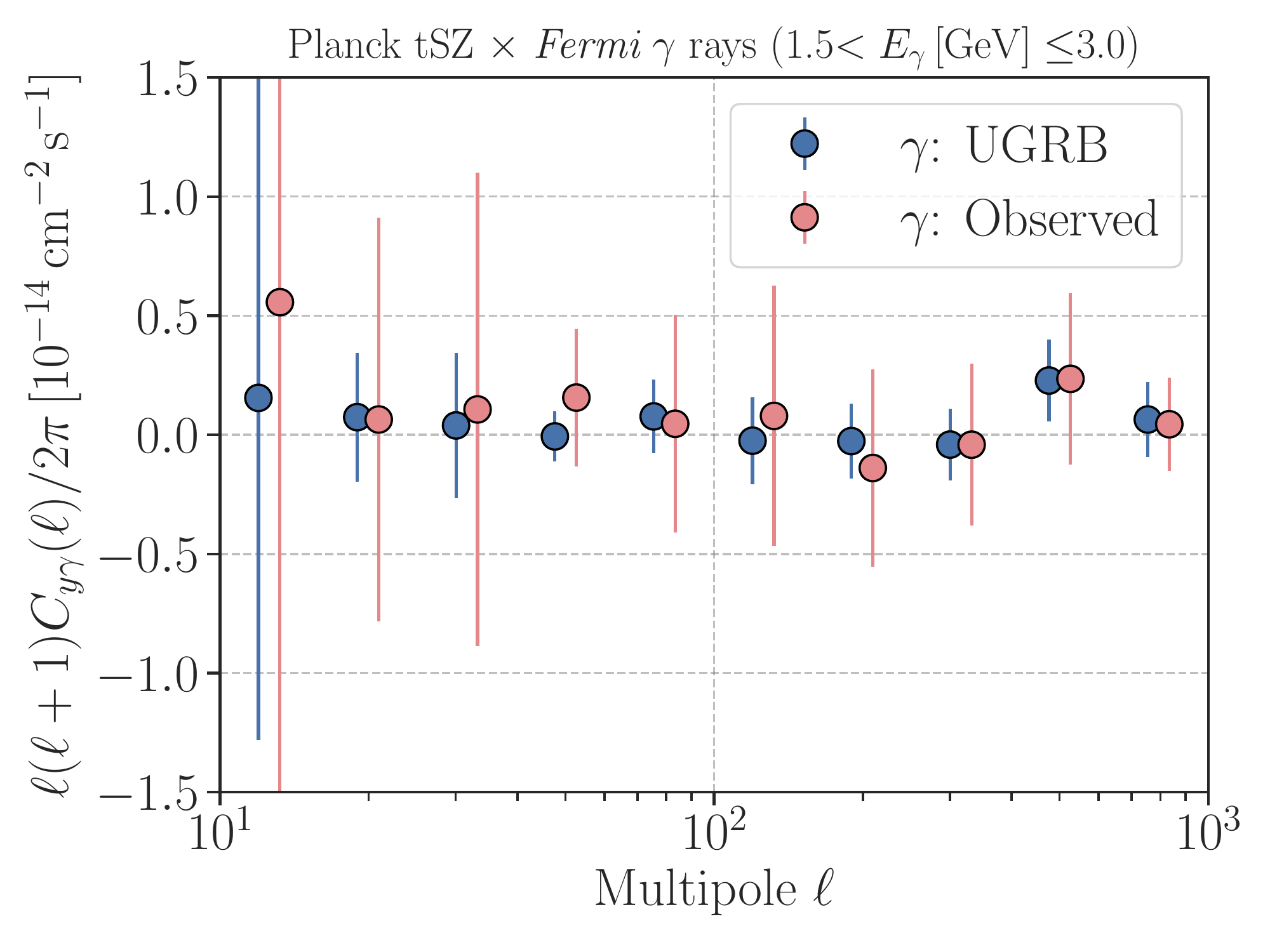}
       \includegraphics[clip, width=1.0\columnwidth]
       {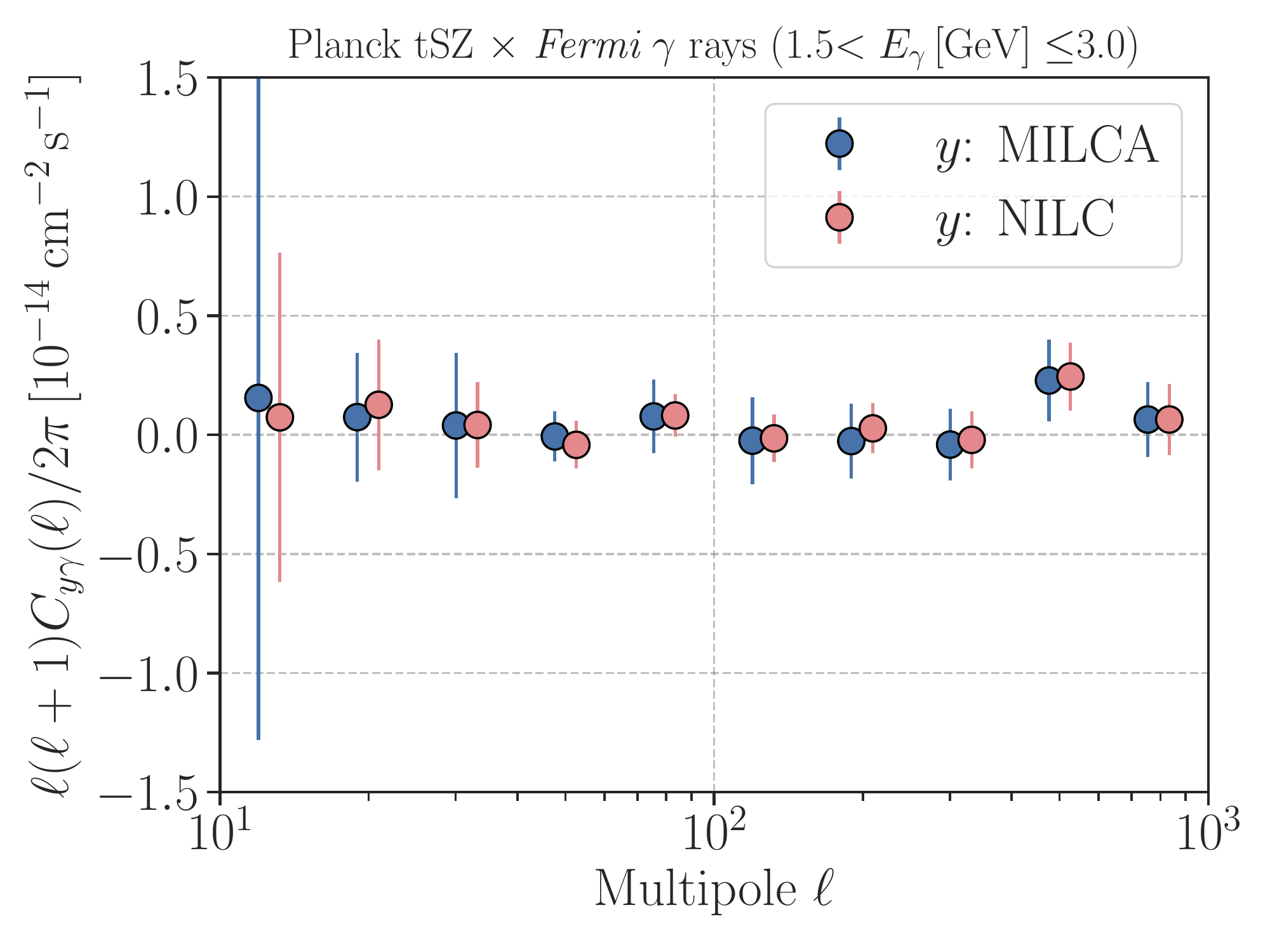}
       \includegraphics[clip, width=1.0\columnwidth]
       {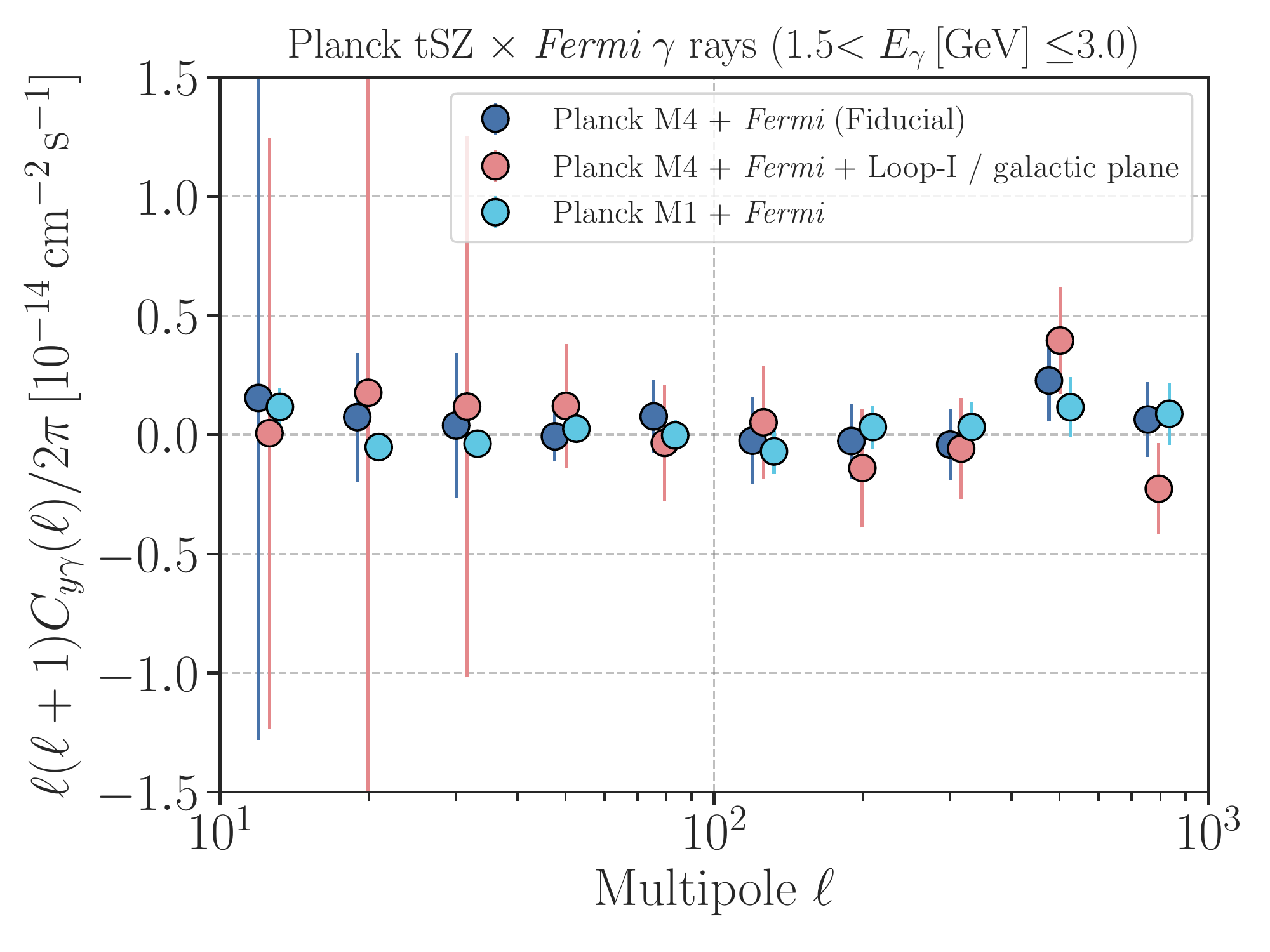}
     \caption{
     \label{fig:test_cl_diffg_diffy_mask}
     Dependence of the power spectrum measurement on 
     subtraction of Galactic gamma-ray components,
     difference in component separation methods in CMB,
     and details of masks.
     {\it Top left}: The impact of inaccurate subtraction of Galactic components in the gamma-ray data. The blue points show our fiducial analysis, while the red ones are for the analysis without the subtraction of Galactic gamma rays.
     {\it Top right}: The impact of the detail of the component separation in the CMB data. The blue points show our fiducial analysis, while the red points are the results based on the tSZ map based on another approach.
     {\it Bottom}: The masking effect of the power spectrum analysis. The blue shows the fiducial case, 
     while the red points show the results with masking the Fermi Bubble and Loop-I regions as well as the conservative mask around the Galactic plane. The cyan points represent the most aggressive analysis based on the 40\% Galactic and point-source masks in the CMB.
  } 
    \end{center}
\end{figure*}

In this appendix, we investigate some systematic 
uncertainties in the measurement of the UGRB-tSZ power spectrum. We examine three analyses below:

\begin{enumerate}

\item[(A)] We perform the cross-correlation analysis by using the observed gamma-ray intensity. This analysis can validate the effect of the subtraction of Galactic gamma rays in the power spectrum analysis.

\item[(B)] We measure the power spectrum with the UGRB map
and the tSZ map based on the \texttt{NILC} method. This analysis will be useful to check 
if our measurement is sensitive to the detail of the component separation in the CMB.

\item[(C)] We measure the power spectrum with the UGRB map and the fiducial tSZ map (based on the \texttt{MILCA} method), 
but we change the masked regions. We examine three cases of masking: (C1) our fiducial mask,
(C2) the 60\% Galactic/point source mask in the CMB and the masking around the gamma-ray sources, the Fermi Bubble and Loop-I regions with a conservative mask of $|b|<30^{\circ}$ about the Galactic plane, 
and
(C3) the 40\% Galactic/point source mask in the CMB and the masking around the gamma-ray sources.
On the mask (C2), we apply a Galactic longitude cut with $0^{\circ} < \ell < 50^{\circ}$ and $260^{\circ} < \ell < 360^{\circ}$ to exclude the Fermi Bubble and Loop-I regions.
The mask (C3) would lead to the most aggressive analysis with the largest sky coverage, but it will be most 
affected by the contamination due to any point sources
or/and the large-scale residual Galactic emission.

\end{enumerate}

Figure~\ref{fig:test_cl_diffg_diffy_mask} 
summarizes the results of our systematic test.
The left top panel shows the analysis 
testing the impact of Galactic gamma rays (case A),
the right top panel represents the effect of the detail in the component separation in the microwave data (case B),
and the bottom panel highlights the masking effect on the power spectrum analysis (case C).
These analyses indicate that our measurement of the power spectrum at $10<\ell<1000$ is less affected by systematic uncertainties due to the imperfect estimates 
of Galactic gamma rays and the tSZ effect, 
the residual contribution from astrophysical sources,
and a possible large-scale correlation between gamma-ray
and microwave observations.


\bibliography{apssamp}

\end{document}